\documentclass[aps,prx,onecolumn,superscriptaddress]{revtex4-2}  

\usepackage{natbib}
\usepackage{graphicx}
\usepackage{amsmath}
\usepackage[inline]{enumitem}
\usepackage{textcomp} 
\usepackage{gensymb} 
\usepackage[mathscr]{euscript}

\usepackage{setspace} 

\usepackage{xr} 

\usepackage{graphicx}
\usepackage{times,ulem,color,bm}
\usepackage{amsmath,amsfonts,amssymb}
\usepackage{ragged2e} 

\usepackage{dutchcal}
\newcommand{\alex}[1]{\textcolor{black}{#1}}
\newcommand{\arn}[1]{\textcolor{black}{#1}}
\newcommand{\mich}[1]{\textcolor{black}{#1}}
\newcommand{\al}[1]{\textcolor{black}{#1}}
\newcommand{\rita}[1]{\textcolor{black}{#1}}
\newcommand{\all}[1]{\textcolor{black}{#1}}

\def\rout{\mathbf{r}_\textrm{out}}
\def\rin{\mathbf{r}_\textrm{in}}
\def\xout{x_\mathrm{out}}
\def\xin{x_\mathrm{in}}
\def\yout{y_\mathrm{out}}
\def\yin{y_\mathrm{in}}

\def\kout{\mathbf{k_\mathrm{out}}}
\def\kpout{\mathbf{k'_\mathrm{out}}}

\def\kin{\mathbf{k_\mathrm{in}}}

\def\Rout{\mathbf{R}_\textrm{out}}

\def\Dout{\mathbf{D}_\textrm{out}}
\def\Doutp{\mathbf{D}_\textrm{out}^{(p)}}
\def\Coutp{\mathbf{C}_\textrm{out}^{(p)}}

\def\Cout{\mathbf{C}_\textrm{out}}

\def\Din{\mathbf{D}_\textrm{in}}

\def\R{\mathbf{R}}
\def\r{\mathbf{r}}

\begin{document}

\title{A Distortion Matrix Framework for High-Resolution Passive Seismic 3D Imaging: Application to the San Jacinto Fault Zone, California}
\author{Rita Touma}
\affiliation{ISTerre, Universit\'{e} Grenoble Alpes, Maison des G\'{e}osciences, BP 53, F-38041 Grenoble, France}
	\affiliation{Institut Langevin, ESPCI Paris, PSL University, CNRS, Univ Paris Diderot, Sorbonne Paris Cit\'{e}, 1 rue Jussieu, F-75005 Paris, France}
\author{Thibaud Blondel}
\affiliation{Institut Langevin, ESPCI Paris, PSL University, CNRS, Univ Paris Diderot, Sorbonne Paris Cit\'{e}, 1 rue Jussieu, F-75005 Paris, France}
\author{Arnaud Derode}
	\affiliation{Institut Langevin, ESPCI Paris, PSL University, CNRS, Univ Paris Diderot, Sorbonne Paris Cit\'{e}, 1 rue Jussieu, F-75005 Paris, France}
\author{Michel Campillo}
\affiliation{ISTerre, Universit\'{e} Grenoble Alpes, Maison des G\'{e}osciences, BP 53, F-38041 Grenoble, France}
\author{{Alexandre Aubry}}
\email{alexandre.aubry@espci.fr}
\affiliation{Institut Langevin, ESPCI Paris, PSL University, CNRS, Univ Paris Diderot, Sorbonne Paris Cit\'{e}, 1 rue Jussieu, F-75005 Paris, France}

\date{\today}


\begin{abstract}
Reflection seismic imaging usually suffers from a loss of resolution and contrast because of the fluctuations of the wave velocities in the Earth's crust. In the literature, phase distortion issues are generally circumvented by means of a background wave velocity model. However, it requires a prior tomography of the wave velocity distribution in the medium, which is often not possible, especially in depth. In this paper, a matrix approach of seismic imaging is developed to retrieve a three-dimensional image of the subsoil, \arn{despite a rough knowledge of the background wave velocity}. To do so, passive noise cross-correlations between geophones of a seismic array are investigated under a matrix formalism. \arn{They form a reflection matrix that contains all the information available on the medium. A set of matrix operations can then be applied in order to extract the relevant information as a function of the problem considered. On the one hand, the background seismic wave velocity can be estimated and its fluctuations quantified by projecting the reflection matrix in a focused basis. It consists in investigating the response between virtual sources and detectors synthesized at any point in the medium. The minimization of their cross-talk can then be used as a guide star for approaching the actual wave velocity distribution. On the other hand, the detrimental effect of wave velocity fluctuations on imaging is overcome by introducing a novel mathematical object: The distortion matrix.}
This operator essentially connects any virtual source inside the medium with
the distortion that a wavefront, emitted from that point, experiences
due to heterogeneities. A time reversal analysis of the distortion matrix enables the estimation of the transmission matrix that links each \textit{real} geophone at the surface and each virtual geophone in depth. Phase distortions can then be compensated for any point of the underground. Applied to \alex{passive} seismic data recorded along the Clark branch of the San Jacinto fault zone, the present method is shown to provide an image of the fault until a depth of 4 km over the frequency range 10-20 Hz  with \alex{an horizontal} resolution of 80 m. Strikingly, this resolution is almost one eighth below the diffraction limit imposed by the geophone array aperture. The heterogeneities of the subsoil play the role of a {scattering lens} and of {a transverse wave guide} which increase drastically the array aperture. The contrast is also optimized since most of the incoherent noise is eliminated by the iterative time reversal process.  Beyond the specific case of the San Jacinto Fault Zone, the reported approach can be applied to any scales and areas for which a reflection matrix is available at a spatial sampling satisfying the Nyquist criterion.
\end{abstract}

\maketitle

\section{Introduction}

Waves constitute a powerful means to non destructively probe an unknown medium. Indeed, wave propagation is fully determined by the wave equation and boundary conditions. As stated by diffraction theory \mich{in the acoustic approximation}, knowing the incident wave-field and the internal properties in terms of density $\rho$ and celerity $c$ theoretically allows to compute the wave field everywhere and at any time inside the medium. \arn{This is the so-called "forward" or "modeling" problem. Conversely, when the medium is unknown, the lack of celerity and density knowledge makes it impossible to compute the spatio-temporal evolution of the wave field.} Yet this evolution can be known, at least at the boundary, through experimental measurement of the wave-field scattered by the medium. The inverse problem then consists in deducing the medium internal properties from the recording of the wave field at its surface. A first way to do so is to assume a velocity and density background, solve the forward problem to compute the time-dependent signal that would be backscattered if the background model was true, \arn{and iteratively update this model to minimize the difference with the actual recordings}. Another way is to directly back-propagate the scattered echoes to reflectors inside the medium. This also amounts to \arn{updating} a background model since a reflector is nothing else than a variation in acoustic impedance $\rho c$. In both strategies, referred to as "inversion" and "migration", respectively, a celerity macro model is required and the purpose is to compute variations from this model under the assumption that they are small (Born approximation). If they are not, the reflected wave-field may be subject to aberrations and multiple scattering that the macro model fails at modeling. These issues lead to distorted images, lack of resolution and unphysical features, which are very detrimental to the imaging process.

In seismic exploration, these issues are important because in most cases the celerity is non constant \arn{in space} and its distribution is unknown. Most geological settings actually consist of several layers of rocks and sediments with distinct mechanical properties as well as location-dependent thickness, faulting and strata organization. These may be difficult to estimate without previous geologic expertise of the subsurface, especially in areas with high lateral mechanical stress that bend and break the layers and make them superimpose. When trying to retrieve details at \alex{a diffraction-limited resolution}, the previous knowledge required to build reliable images may be already fairly demanding. On the one hand, the inversion problem cannot be solved if the initial celerity model is too far from the reality \arn{because the regularization procedure can end up stuck in a local minimum}. On the other hand, migration techniques would lead to loss of resolution due to phase distortions and a blurred image due to multiple scattering. That being said, the question that naturally arises and which the present work aims at addressing is: how to retrieve an accurate image when little to no previous knowledge \arn{on the spatial variations of the wave speed} is available?

To cope with this issue, our strategy is to develop a matrix approach of seismic wave imaging. In a linear scattering medium, a reflection matrix relates the input and output on a single side of the medium. It contains all the relevant information on the medium as it fully describes wave propagation inside the scattering medium. In the last decade, the advent of multi-element arrays with controllable emitters and receivers has opened up a route towards the ability of measuring a reflection matrix \arn{in the case where the input and outputs points are located on the same side of the scattering medium. In particular, the reflection matrix has been shown to be of great interest for detection and imaging purposes in scattering media,} whether it \arn{be} in acoustics~\citep{Robert2008,aubry,shahjahan} or optics~\citep{Kang2015,badon2016}. The reflection matrix contains the set of inter-element impulse responses recorded between each array element. It has already been shown to be a powerful tool for focusing in multi-target media~\citep{prada,Popoff2011}, as well as for separating single and multiple scattering~\citep{aubry2,badon2016} in strongly scattering media. These matrix methods have been successfully applied to geophysics in the extremely challenging case of the Erebus volcano in Antarctica~\citep{erebus}. 
\al{Seismic data redatuming leads to the synthesis of a focused reflection matrix $\mathbf{R}$ containing the impulse responses between a set of virtual geophones mapping the underground to be imaged. This matrix is of particular interest for imaging: its confocal component i.e., diagonal (or close to diagonal) elements result from single scattering, whereas multiple scattering is responsible for a spreading outside the diagonal. By applying an adaptive confocal filter and iterative time reversal to the redatumed data, most of the multiple scattering background is removed, thereby revealing main internal structures of the volcano. Yet, the resulting ``confocal'' image still suffers from the phase
distortions induced by the long-scale fluctuations of the seismic velocity. In the present paper, we show that the off-diagonal coefficients of $\mathbf{R}$ can be taken advantage of instead of being tossed away. Phase distortions are overcome, which improves the confocal image of the subsoil with a diffraction-limited transverse resolution.}


\alex{To that aim, we will rely on the \textit{distortion matrix} concept that} has been recently introduced by two seminal works in ultrasound imaging~\citep{william} and in optical microscopy~\citep{Badon2019}.  Inspired by the pioneering work of~\cite{Robert2008}, the distortion matrix $\mathbf{D}$ is defined between a set of incident plane waves~\citep{Montaldo2009} and the set of \alex{virtual geophones} inside the medium~\citep{Robert2008}. It contains the \textit{deviations} from an ideal reflected wavefront which would be obtained in the absence of inhomogeneities.  As shown by recent studies~\citep{Badon2019,william}, a time reversal analysis of the $\mathbf{D}$-matrix allows to synthesize virtual reflectors in depth. This process can then be leveraged for unscrambling the phase distortions undergone by the incident and reflected wavefronts.  \alex{This matrix imaging approach has been shown to be particularly robust since it applies to all kind of scattering regimes: point-like targets~\citep{william}, specular reflectors~\citep{Badon2019} or randomly distributed scatterers~\citep{william} etc. The range of wave velocity distributions on which this matrix method can be applied is \al{vast}. It goes from the simple case of multi-layered media~\citep{william} to strongly heterogeneous media~\citep{william,Badon2019} displaying both lateral and axial variations of the wave velocity. These proof-of-concept studies made on both synthetic samples, \textit{ex-vivo} or \textit{in-vivo} tissues have thus shown the \al{power} of the matrix imaging approach to solve very different imaging problems whatever the nature of waves and length scales.  \al{It only requires a} spatial sampling of the recorded wave-field that meets the Nyquist criterion. The present paper aims at demonstrating the relevance of this matrix imaging approach for geophysical imaging.} 

\alex{Indeed, }overcoming phase distortions \alex{induced by wave velocity variations} would be especially valuable for geophysical applications given the stratified structures of the environments of interest. Migration techniques in Fourier domain have actually been very popular for imaging in layered media \citep{stolt_fk,gazdag_phaseshift}, however they only hold for 1D celerity models with no lateral variations. Subsequent works have focused on adapting these techniques to take into account increasing lateral velocity variations, at the cost of more numerical and computational complexity \citep{gazdag_phaseinterp,iei,biondi}. Contrary to these well-established methods, the matrix approach does not require any assumption on the structures and on the velocity distribution inside the medium, while being fairly light on the computational aspect. The present paper aims at studying the relevance of the matrix approach for geophysical imaging. 
 
\arn{Coherent sources (vibrating trucks, explosives, \textit{etc.}) can be used in shallow subsurface ($<$1 km) imaging. Incoherent signals (seismic noise) can also be taken advantage of for imaging purposes. It was shown,}
 twenty years ago, how a coherent information can be extracted from this incoherent seismic noise. Under appropriate wave-field conditions, the cross-correlation of seismic noise recorded by two stations was actually shown to yield the \arn{impulse response} between them~\citep{weaver,campillo,derode2,wapenaar,snieder2,larose}, providing new opportunities to develop imaging techniques without using active sources. As surface waves dominate ambient noise, \arn{most papers on the topic} aimed at extracting surface wave properties from ambient noise correlations~\citep{shapiro,sabra2005,yang2007}. However, a few studies also reported the retrieval of body wave reflection from noise correlations~\citep{roux2005,draganov2007,draganov2009,poli2012a}. Reflected body waves contain information about the subsurface and allow the imaging of deep structures with an improved resolution~\citep{ruigrok2010}. {Strikingly,} \cite{poli2012b} showed the possibility of mapping the upper mantle discontinuities (at 410 and 660 km of depth) by extracting body waves reflection from ambient noise\mich{, while \cite{Retailleau2020} mapped a region of the core-mantle boundary at about 2900 km depth.}
 
 In this paper, inspired by the reflection matrix approach developed  by~\cite{erebus} and based on noise cross-correlations, the distortion matrix approach is extended to satisfy seismic imaging purposes. \all{Compared to previous works~\citep{Badon2019,Lambert2020}, the matrix imaging method is here refined to take the best advantage of the nature of scatterers in geophysics (sparse scattering). } The method \all{thus developed} is applied to San Jacinto fault zone (SJFZ) site. Fault zones are indeed among the most challenging media for seismic imaging {given their} highly localized and abrupt variations of mechanical properties, extensive fractures and damage zones. In that respect, the {SJFZ} is the most seismically active fault zone in Southern California~\citep{Hauksson}{. It} accounts for a large portion of the plate motion in the region~\citep{Johnson1994,Lindsey2013}. 
 A highly complex fault-zone structure with prominent lateral and vertical heterogeneities at various scales have already been highlighted in previous studies~\citep{Allam2012,Zigone2014,Roux_2016}. In particular, maps of the P and S wave velocities{, $V_P$ and $V_S$,} have been inverted from earthquake arrival times for a depth range of 2-20 km
~\citep{Allam2012,allam2014seismic}. {Surface wave tomographic images built from noise correlations revealed the velocity structure in the top 7 km of the complex plate boundary region at a resolution of about ten kilometers ~\citep{Zigone2014}}. To complement these regional studies and provide structural features in the first few kilometers with an improved resolution, ambient noise at higher frequency up to 10 Hz was analysed from data recorded by a dense rectangular array deployed around the Clark branch of the SJFZ ~\citep{Ben_Zion_2015, Roux_2016, mordret2019}.
{In particular}, ~\cite{zigone2019imaging} used ambient noise cross-correlations in the 2-35 Hz frequency range to derive a velocity model in the top 100 m with a resolution of 50 m.

{Imaging deeper the fault area at such resolution is challenging because of the damage and the complex distribution} of small-scale heterogeneities. {Yet, a much larger penetration depth can be expected by taking advantage of the reflected bulk waves.} To do so, the matrix approach of seismic imaging is particularly useful since it only \arn{requires a rough idea of the mean wave velocity}. Besides, it shall provide a three-dimensional image of the subsoil acoustic impedance instead of just the wave velocity. To implement this matrix approach, we take advantage of a spatially dense array of geophones deployed over the damage zone of SJFZ~\citep{Ben_Zion_2015}. Noise cross-correlations are used to retrieve the \arn{impulse responses} between the geophones\alex{~\citep{Ben_Zion_2015,Roux_2016}}.
 The associated passive reflection matrix is then investigated to image the first few kilometers of the crust by virtue of body waves emerging from
noise correlations. As a whole, the process we present in this paper can be analyzed as a combination of six building blocks: 
\begin{itemize}
\item (B1) A {Fourier transform} of the recorded signals yields a set of response matrices $\mathbf{K}(f)$ associated with the dense array of geophones.
\item (B2) Based on a rough estimate of velocity $c_0$, a double focusing operation is performed both at emission and reception by means of simple matrix operations. A set of focused reflection matrices $\mathbf{R}(f,z)$ are obtained at any arbitrary depth $z$ below the surface.
\item (B3) A coherent sum of these matrices over the frequency bandwidth yields a broadband reflection matrix $\mathbf{R}(z)$ at any depth $z$.
\item (B4)  By projecting the input or output entries of this matrix in the far-field, the distorted component $\mathbf{D}$ of the reflected wave-field can be extracted.
\item (B5) A virtual iterative time reversal process is applied to the matrix $\mathbf{D}$ to extract the phase distortions undergone by the incident or reflected wave-fields during their travel from the Earth surface to the focal plane.
\item (B6) The whole process converge towards the focusing laws that shall be applied at input or output of the reflection matrix in order to compensate for aberrations.
\end{itemize}
As a result of these six steps, an in-depth confocal image of the SJFZ is built. While conventional migration methods lead to a badly resolved image of the SJFZ subsoil, the matrix approach clearly reveals sedimentary layers close to the surface \arn{($z<1000$ m)} and several geological layers at larger depth \arn{($1000$ m$<z<4000$ m)}. The layers structure is shown to be different on each side of the fault. Large dip angles are also highlighted in the vicinity of the fault. A structural interpretation of the obtained images can be finally built on the existing literature about SJFZ. 
 
\section{Reflection matrix}%
\begin{figure*}
 \centering
 \includegraphics[width=14cm]{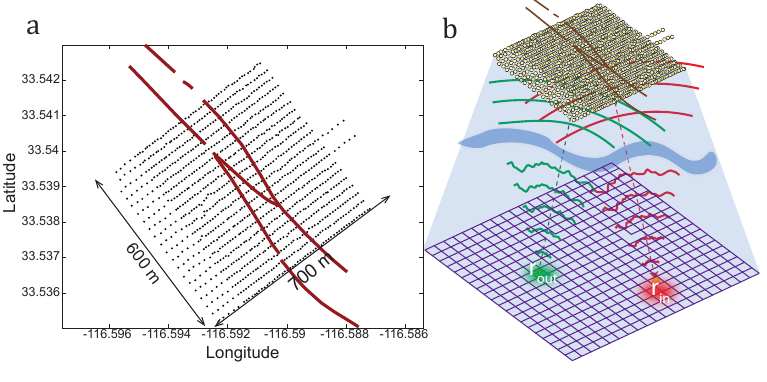}
 \caption{(a) Map of the 1108 (10 Hz) geophones installed in a \arn{600 m $\times$ 700 m} configuration above the Clark branch (red lines) of the San Jacinto Fault (Southern California). Each row along the x-direction is composed of $\sim$55 sensors with a pitch of 10 m, and the nominal separation between the rows in the $y$-direction is 30 m. Seismic ambient noise was recorded over more than one month, in May-June 2014. (b) Adaptive focusing at emission and reception on two points $\rin$ and $\rout$ of the focal plane ($z = c_0t/2$) yields the \arn{impulse response} between virtual geophones placed at these two points. The same operation is repeated for any couple of points in the focal plane and yields the focused reflection matrix $\R$.}
 \label{fig1}
  \end{figure*}

\subsection{Response matrix between geophones}

 The data used in this study has been measured from May 7, 2014 to June 13, 2014 by a spatially dense Nodal array consisting of 1108 vertical geophones straddling the Clark Branch of SJFZ, southeast of Anza~\citep{Ben_Zion_2015}. Figure \ref{fig1}a shows the location of the 1108 vertical geophones organized as a \arn{600 m$\times$700 m} grid with inter-station distances \arn{ $\delta u_x\sim$ 10m and $\delta u _y\sim$ 30m}. This array has been continuously recording the ambient noise at 500 sample.s$^{-1}$, from which cross-correlation has been performed after whitening in the 10-20 Hz range with time lags ranging from $-5$ s to $+5$ s. This provides an estimate of the impulse response between every pair of geophones. Each geophone is denoted by an index $i$ and its position $\mathbf{s_i}$. The impulse response between stations $i$ and $j$ is noted $k_{ij}(t)$, with $t$ the time lag. The set of impulse responses \alex{forms} a time-dependent response matrix $\mathbf{K}(t)$. 
 
 Given the high density of the network, \arn{neighbouring} geophones belong to the same coherence area of seismic noise. The characteristic dimension of this area is indeed of ${\lambda}/{2}\sim$ 50 m which is larger than the interstation distance \arn{$\delta u$}. This is responsible for a strong auto-correlation signal around $t=0$ for geophones located in the same coherence area. This \arn{peak} is proportional to the seismic noise power and does not account for the impulse response between neighbour geophones. To \arn{prevent this artifact from spoiling} the subsequent analysis, a prior filter has been applied to the data in order to reduce the weight of the corresponding impulse responses $k_{ij}(t)$ whose associated geophones $i$ and $j$ are contained in the same coherence area {(see \all{Supplementary Section~S1)}}. 

{The impulse responses exhibit several direct arrivals that have already been investigated by \cite{Ben_Zion_2015} and \cite{Roux_2016}. Ballistic waves, likely direct inter-station S-wave and P-wave, arrive before the Rayleigh wave at apparent velocities larger than 1000 m/s.}
\cite{Roux_2016} used iterative double beamforming to map the phase and group velocities {of Rayleigh waves across the fault in the 1-5 Hz frequency bandwidth.} Subsequently, \cite{mordret2019} inverted these dispersion curves to build a 3-D shear wave velocity model around the Clark fault down to 500 m depth. \arn{Assuming the  $V_p/V_s$ ratio to be a linear function of depth,} the following averaged value was found for the P-wave velocity over the top 800 m: $V_p\sim $1500 m/s. More recently, the P-wave velocity distribution in the 100 m-thick shallow layer has also been inverted using travel time data associated with active shots~\citep{share2020}. Low-velocity structures were detected, associated with a shallow sedimentary basin ~\citep{hillers2016, mordret2019, share2020} and a fault zone trapping structure ~\citep{Ben_Zion_2015,qin2018}.

{To the best of our knowledge, an accurate model of $V_p$ in the SJFZ region is not available beyond this shallow layer.} As a consequence, in the present study, we will use an approximated homogeneous P-wave velocity model of $c_0= 1500$ m/s. This choice will be validated and~ discussed \textit{a posteriori} by a minimization of the aberration effects in the 3D image \alex{(see Supplementary Fig.~S3)}. We are not interested in the ballistic component of the wave-field {but rather in} its scattered contribution due to reflections by the in-depth structure along the fault. {\al{The beamformed echoes used in our matrix imaging process} are mainly associated with P-waves since only the vertical component of the \al{impulse responses between geophones} is considered in this study.} Unlike our previous study on the Erebus volcano~\citep{erebus}, the scattered wave-field {consists of} a single scattering contribution which is \textit{a priori} largely predominant compared to the multiple scattering background. This will be confirmed \textit{a posteriori} by the reflection matrix features (see Sec.~\ref{focal}). Singly-scattered echoes can then be taken advantage of to build a 3D image of the subsoil reflectivity. This local information can be retrieved from $\mathbf{K}(t)$ by applying appropriate time delays to perform focusing in post-processing, both in emission and reception. \arn{While focusing in emission consists in applying proper time delays in the recorded seismic data so that they constructively interfere at an arbitrary position at depth, focusing in reception consists in applying proper time delays in the recorded seismic data so that the information coming from an arbitrary position at depth constructively interfere.}
Based on the Kirchoff-Helmholtz integral, such a focusing operation is standard in exploration seismology and referred to as \textit{redatuming}~\citep{Berkhout,Berryhill,Berkhout2}. However, in the present case, the strongly heterogeneous distribution of the seismic wave velocities induces strong phase distortions that degrade this imaging process. A prior quantification and correction of these phase distortions is thus required to reach a diffraction-limited \alex{lateral} resolution and an optimized contrast for the image. As we will see, a matrix formalism is \arn{a well-matched} tool to locally capture such \arn{information.}

\subsection{\label{focal} Focused reflection matrix}

The reflection matrix can be defined in general as \arn{an ensemble of responses, each response linking one vector to another vector. The type of vector coordinates will be referred to as \textit{bases}. They can be spatial coordinates (hence the vector refers to an actual point within or at the surface of the medium,  see Fig.~\ref{fig1}b) or wave vector coordinates. Various bases are involved in this work :} (\textit{i}) the recording basis ($\mathbf{u}$)\arn{, whose elements are the positions of the geophones}\alex{;} (\textit{ii}) the focused basis ($\mathbf{r}$) \arn{which corresponds to the positions of  \textit{virtual geophones} at which focusing at emission or reception is intended;} and (\textit{iii}) the Fourier basis ({$\mathbf{k}$}). \arn{Because of linearity and time-invariance, seismic data can be  projected from the recording basis to the focused basis by a simple matrix product.} In the frequency domain, simple matrix products allow seismic data to be easily projected from the recording basis to the focused basis where local information on the medium properties can be extracted~\citep{badon2016,erebus,Lambert2020}. 

Consequently, {we} first apply a temporal Fourier transform to the response matrix to obtain a set of monochromatic matrices $\mathbf{K}(f)$. To project $\mathbf{K}(f)$ into the focused basis, we then define a free-space Green's matrix,  $\mathbf{G_0}(f)$, which describes the propagation of waves between the geophones and focused basis. Its elements correspond to the \alex{causal} 3D Green's functions which connect the geophone's transverse position $\mathbf{u}$ to any focal point defined by its transverse position $\mathbf{r}$ and depth $z$ in a supposed homogeneous medium:
\begin{equation}
{G}_{0}(\mathbf{r},\mathbf{u},z,f) = \frac{\textrm{e}^{-j 2\pi f \sqrt{\left\|\mathbf{r}-\mathbf{u}\right\|^2+z^2}/c_0}}{4 \pi \sqrt{\left\|\mathbf{r}-\mathbf{u}\right\|^2+z^2}}
\label{Gfreespace}
\end{equation}
$\mathbf{K}(f)$ can now be projected both in emission and reception to the focused basis via the \arn{following matrix product at each depth~$z$}~\citep{erebus,Lambert2020}:
\begin{equation}
\label{projRrr}
\R(z,f)=\mathbf{{G}}_0^*\left(z,f\right) \times \mathbf{K}(f)\times \mathbf{G}_0^{\dagger}\left(z,f \right), \end{equation}
where the symbols $*$, $\dagger$ and $\times$ stands for phase conjugate, transpose conjugate and matrix product, respectively. Equation (\ref{projRrr}) simulates focused beamforming in post-processing in both emission and reception. \alex{The choice of causal Green's function in the propagation matrix $\mathbf{G}_0$ (Eq.~\ref{Gfreespace}) implies that beamformed echoes are associated with down-going waves at the input and up-going waves at the output (see Fig.~\ref{fig1}b).} \arn{Each coefficient of the focused reflection matrix $\mathbf{R}(z,f)$ involves pairs of virtual geophones,} $\rin=(\xin,\yin)$ and $\rout=(\xout,\yout)$, which are located at the same depth $z$ (Fig.\ \ref{fig1}b). For broadband signals, ballistic \arn{time-gating} can be performed to select only the echoes arriving at the ballistic time $t_B$ in the focused basis~\citep{Lambert2020}: \arn{$t_B={(\left\|\mathbf{r}_\textrm{in}-\mathbf{u}_\textrm{in}\right\|+\left\|\mathbf{r}_\textrm{out}-\mathbf{u}_\textrm{out}\right\|)/c_0}$.  Under a matrix formalism, this time-gating can be performed by means of a coherent sum of $\R(f)$ over the frequency bandwidth \alex{$\Delta f=10$ Hz}. It yields the broadband} focused reflection matrix
\begin{equation}
\label{Rrr_broadband}
\R (z) = \int^{f{+}}_{f_{-}} df \R (z,f),
\end{equation} 
where \arn{$f_{\pm}=f_0 \pm \Delta f/2$} and \alex{$f_0=15$ Hz} is the central frequency. Each element of $\R(z)$ contains the \arn{complex amplitude} of the wave that would be detected by a virtual detector located at $\rout=(\xout,\yout)$ just after a virtual emitter at $\rin=(\xin,\yin)$ emits a brief pulse of length \arn{$\delta t= \Delta f^{-1}$} at the central frequency $f_0$. Importantly, the broadband focused reflection matrix synthesizes the responses between virtual geophones which have a greatly reduced axial dimension \arn{$\delta z=c \delta t$ compared to their stretching $\delta z_0=2 \lambda/\sin^2 \theta$ in the monochromatic regime~\citep{born}. $\theta=\arctan (\mathcal{D} /2z)$ is the maximum angle under which the geophone array is seen from the common mid-point and \arn{$\mathcal{D} \sim 700$ m}, the characteristic size of the geophone array. As a consequence, considering a broadband reflection matrix $\R (z)$} will significantly improve the \alex{vertical} resolution of the subsequent analysis. For the sake of a lighter notation, we will omit, in the following, the dependence in $z$ but keep in mind that the focused reflection matrix differs at each depth.
\begin{figure*}
 \centering
 \includegraphics[width=14cm]{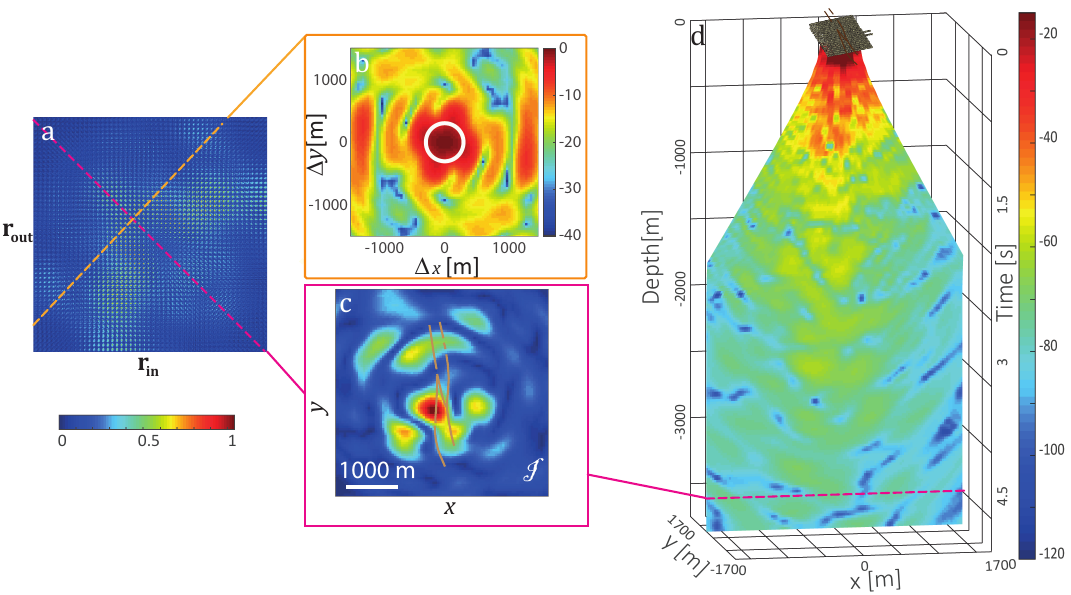}
 \caption{(a) Focused reflection matrix $\mathbf{R}$ in the focal plane at depth $z=$3600 m.
(b) Imaging PSF deduced from the antidiagonal of $\mathbf{R}$ in (a) whose common mid-point exhibits the maximum confocal signal. The white circle accounts for the theoretical \alex{transverse} resolution cell imposed by the geophone array aperture. (c) Confocal image extracted from the diagonal of $\mathbf{R}$ in (a). (d) Vertical slice of the 3D confocal image obtained by \arn{combining} the diagonal of $\R$ at each depth. The slice orientation is chosen to be normal to the fault. {The color scale on the bottom left is linear and applies to panels (a,c). \alex{The color scale in panels (b,d) is in dB.}} }
\label{fig2}
\end{figure*}

Figure~\ref{fig2}a displays one example of the broadband focused reflection matrix $\R$ at depth $z=$3600 m. \arn{
In the case of SJFZ, it appears that \alex{a part} of the backscattered energy is still concentrated in the vicinity of the diagonal of the focused reflection matrix at $z=3600$ m (Fig.~\ref{fig2}a); this is very different from the Erebus volcano for which the reflection matrix displayed a fully random feature~\citep{erebus}.} This indicates that single scattering dominates at this depth\arn{: The beam is focused, scattered just once, and focused in reception. On the contrary, a broadening of the back-scattered energy outside the diagonal would mean that the beam undergoes aberration and/or  multiple scattering. In fact, the diagonal elements of\ $\R$ {($\rin=\rout$)} correspond to what would be obtained from confocal imaging: transmit and receive focusing are simultaneously performed on each point in the medium.} A confocal image can thus be \arn{obtained} from the diagonal elements of $\R$, computed at each depth: 
\begin{equation}
\label{imcalc}
\all{\mathcal{I}}\left(\mathbf{r},z\right)\equiv \left|R\left(\r,\r,z\right)\right|^2.
\end{equation}
Figure \ref{fig2}c \arn{displays} the 2D confocal image built from the diagonal of the reflection matrix in Fig.~\ref{fig2}a at \alex{ballistic time} $\alex{t_B}=4.6$ s, hence at an effective depth $z=c_0\alex{t_B}/2=3600$ m. Some scattering structures seem to arise at different locations along the fault but confocal imaging is extremely sensitive to aberration issues. One thus has to be very careful about the interpretation of a raw confocal image. This observation is confirmed by Fig.~\ref{fig2}d that displays \alex{a cross-sectional view of the SJFZ underground}. Each speckle grain in this image occupies a major part of the field-of-view. Hence, aberrations seem to be pretty intense at large depths (beyond 1500 m) and the inner structure of the SJFZ cannot be deduced from a basic confocal image.

\arn{Fortunately}, the matrix $\R$ contains much more information than a single confocal image. In particular, focusing quality can be assessed by means of the off-diagonal elements of $\R$. To understand why, $\R$ can be expressed theoretically as follows\all{~\citep{Lambert2020,lambert_ieee}}:{
\begin{equation}
\label{RxxMatrix}
\R=\mathbf{H}_\textrm{out}^{\top}\times \mathbf{\Gamma} \times \mathbf{H}_\textrm{in},
\end{equation}}
\arn{where the symbol $\top$ stands for transpose.} The matrix $\mathbf{\Gamma}$ describes the scattering process inside the medium. In the single scattering regime, $\mathbf{\Gamma}(z)$ is diagonal and its coefficients map the local reflectivity $\gamma(\mathbf{r})$ of the subsoil. {$\mathbf{H}_\textrm{out}=[{H}_\textrm{out}(\mathbf{r},\mathbf{r}_\textrm{out})]$ and \all{$\mathbf{H}_\textrm{in}=[{H}_\textrm{in}(\mathbf{r},\mathbf{r}_\textrm{in})]$} are the output and input focusing matrices, respectively.} Their columns correspond to the transmit or receive \mich{point spread functions (PSFs)}, i.e. the spatial amplitude distribution of the focal spots around the focusing point $\rin$ or $\rout$.  For spatially-invariant aberration, \all{we have
\begin{equation}
\label{iso}
 H_\textrm{out/in}(\r,\mathbf{r}_\textrm{out/in})=H_\textrm{out}(\r-\mathbf{r}_\textrm{out/in}). 
\end{equation} 
In that case,} the previous equation can then be rewritten in terms of matrix coefficients as follows:
\begin{equation}
\label{Rrr_coef}
R(\rout,\rin)=\int d\r  H_\textrm{out}(\r-\rout)  \gamma(\r) H_\textrm{in}(\r-\rin) .
\end{equation}
This last equation confirms that the diagonal coefficients of $\R$, i.e. an horizontal slice of the confocal image, result from a convolution between the medium reflectivity $\gamma$ and the product of the input and output PSF{, $H_\textrm{out}\times H_\textrm{in}$}. 

\arn{Interestingly, the off-diagonals terms in the reflection matrix can be exploited to estimate the imaging PSF, and thereby assess the quality of focusing.} To that aim, the relevant observable is the intensity distribution along each antidiagonal of $\R$,{
\begin{eqnarray}
  I(\mathbf{r}_m,\Delta r ) & = & R(\r_m - \Delta \r ,\r_m + \Delta \r) \nonumber \\
  & = & \int d\r'  H_\textrm{out}(\r'-\Delta \r)  \gamma(\r'+\r_m) H_\textrm{in}(\r'+\Delta \r) .
  \label{IrDr_spec}
\end{eqnarray}}
\arn{All couple of points on a given antidiagonal have the same midpoint $\mathbf{r}_m=(\rout+\rin)/2$ , but different spacings $\Delta r= (\rout-\rin)/2$.} Whatever the nature of the scattering medium, the common midpoint intensity profile is a direct indicator of the local PSF. However, its theoretical expression differs slightly depending on the characteristic length scale $l_{\gamma}$ of the reflectivity $\gamma(\mathbf{r})$ at the ballistic depth and the typical width { $\delta_\textrm{in/out}^{(0)}$} of the PSFs~\citep{Lambert2020}. In \alex{layered media} ($l_{\gamma}>>\delta_\textrm{in/out}^{(0)}$),  the common-midpoint amplitude is directly proportional to the convolution between the coherent {output and input PSFs, \arn{$ \left [H_\textrm{out}  \otimes  H_\textrm{in} \right ] (2\Delta \r)$}} (the symbol $\otimes$ stands for convolution). In {the speckle regime ( $l_{\gamma}<<\delta_\textrm{in/out}^{(0)}$)}, the common midpoint intensity $I(\mathbf{r}_m,\Delta \r )$ is directly proportional to the convolution between the incoherent {output and input PSFs, $\left [ |H_\textrm{out}|^2 \otimes |H_\textrm{in}|^2 \right ] (2\Delta \r)$\all{~\citep{lambert_ieee}}.} In the present case, the subsoil of SJFZ can be assumed as a \textit{sparse} scattering medium. It means that only a few bright and coherent reflectors emerge at each depth. This hypothesis will be verified \textit{a posteriori} with the three-dimensional image we will obtain. For an \arn{isolated scatterer}, the common mid-point intensity at \arn{ its position} scales as the product between the two PSFs, \arn{$H_\textrm{out}(\r_m-\Delta \r)\times H_\textrm{in} (\r_m-\Delta \r)$}. Therefore, the energy spreading in the vicinity of each scatterer position shall enable one to probe the spatial extension of the PSF. As the scatterer position is \textit{a priori} unknown, the imaging PSF will be, in practice, probed by considering the antidiagonal whose common mid-point exhibits the maximum confocal signal.

Figure~\ref{fig2}b shows the corresponding common midpoint intensity profile for the matrix $\R$ displayed in Fig.~\ref{fig2}a. It shows a significant spreading of energy over off-diagonal coefficients of $\R$. This effect is a direct manifestation of the aberrations sketched in Fig.~\ref{fig1}b. Indeed, in absence of aberration, all the back-scattered energy would be contained in a diffraction-limited confocal focal spot, $H^2(\Delta \r)=\textrm{sinc}^2 \left ( \pi \Delta r /\delta_0 \right)$, with $\delta_0=\lambda / (2 \sin \theta)$. The ideal -6dB main lobe width (or full width at half maximum) is roughly equal to $\delta_0 \sim 600$ m. This diffraction-limited \alex{lateral} resolution is depicted by a white circle in Fig.~\ref{fig2}b. Here the characteristic size of the main central lobe is $ \delta_\textrm{in/out}^{(0)} \sim  $ 1200 m at $z=3600$ m. Hence, the back-scattered energy spreads well beyond the diffraction limit. \arn{Besides this central lobe, few secondary lobes also emerge in Fig.~\ref{fig2}b due to the gap between the velocity model and the actual seismic wave velocity distribution in SJFZ. \alex{These side lobes are \al{analogous} to cycle skipping effects that occur in full- waveform inversion~\citep{Yao2019}.} As shown in Supplementary Section \alex{S3}, \alex{they} are strongly affected by our choice of $c_0$. Hence this observable can be used for optimizing our wave propagation model. As shown by Supplementary Fig.~\alex{S3}, the value $c_0=1500$ m/s is the seismic wave velocity that clearly minimizes the level of these secondary lobes.  }

\arn{Despite this optimization, the focusing quality remains far from being ideal because of the heterogeneous distribution of $c$ in the subsoil. In the following, we will show how this fundamental issue can become a strength since it can enlarge virtually the aperture angle under which the geophone array is seen, thereby leading to an enhanced \alex{horizontal} resolution.}

\section{Distortion matrix}
 \arn{
To that aim, a new operator is introduced: The so-called distortion matrix {$\mathbf{D}$}~\citep{Badon2019,william}. This operator essentially connects each virtual geophone with the distortion exhibited by the associated wave front in the far-field. The $\mathbf{D}$-matrix is thus equivalent to a reflection matrix but in a moving frame, \textit{i.e} centered around each input focusing beam. This change of frame will allow us to unscramble the contribution of phase aberrations from the medium reflectivity. Last but not least, it will be shown to be particularly efficient for spatially distributed aberrations. While conventional adaptive focusing techniques are only effective over a single isoplanatic patch, \all{the typical area over which aberrations are spatially-invariant,} the $\mathbf{D}$-matrix is an adequate tool to discriminate them and address them independently. }
 \begin{figure*}
 \centering
 \includegraphics[width=17cm]{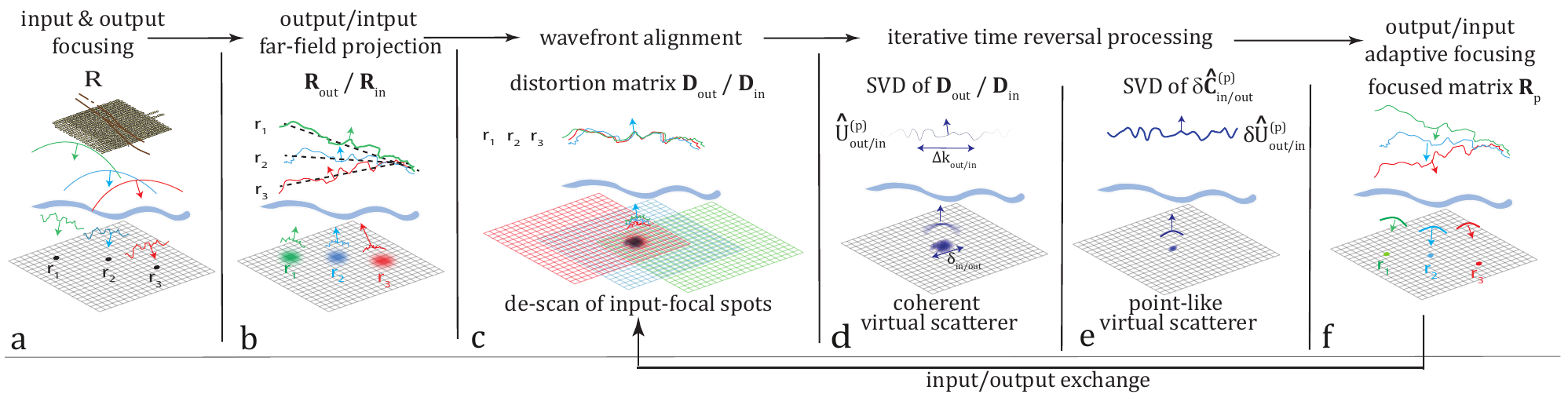}
 \caption{\textbf{Time reversal analysis of the distortion matrix}. {(a) {In the matrix imaging scheme, each point in the focal plane is probed by means of a focused beam at input and output}. (b) A far-field projection of the focused reflection matrix [Eq.~(\ref{Rtt})] yields the dual-basis matrix {$\mathbf{R}$}. (c) By subtracting from each reflected wave-front the geometrical \alex{phase law} \alex{which would be obtained for a perfectly homogeneous medium of wave velocity $c_0$ } [Eq.~(\ref{D_eq})], a distortion matrix {$\mathbf{D}$} is obtained. {$\mathbf{D}$} is equivalent to a reflection matrix but with a static input PSF { $H(\mathbf{r})$} scanned by moving scatterers (Eq.~\ref{Dalex}). (d) {The SVD of the distortion matrix} enables the synthesis of a virtual coherent reflector of scattering distribution {$|H(\mathbf{r})|^2$} \all{(see Supplementary Section S2)} and an estimation of the aberration phase transmittance {$\tilde{H}^{(p)}$ for} each isoplanatic patch $p$ in the field-of-view [Eq.~(\ref{U1})]. {(e) This estimation can be refined by considering, in \all{the second part} of the process, the SVD of the normalized time reversal operator {$\delta \mathbf{\hat{C}}^{(p)}$ \all{\all{(see Supplementary Section S4)}}}. This operation makes the virtual reflector point-like and the estimation of {$\tilde{H}^{(p)}$} more precise. (f) The phase conjugate of {$\tilde{H}^{(p)}$}} yields the focusing law to scan the corresponding isoplanatic patch $p$ and synthesize a novel focused reflection matrix \all{{$\R_p$} [Eq.~\ref{Rpf}]}. } }
\label{fig3}
  \end{figure*}

\subsection{Reflection matrix in a dual basis}

The reflection matrix $\R$ is first projected into the Fourier basis in reception. To that aim, we define a free-space transmission matrix $\mathbf{T}_0$ which corresponds to the Fourier transform operator. Its elements link any transverse component $\mathbf{k}=(k_x,k_y)$ of the wave vector in the Fourier space to the transverse coordinate $\r=(x,y)$ of any point in an ideal homogeneous medium:
\begin{equation}
\label{T0}
T_0\left(\mathbf{k},\r \right) =\exp{\left(i \mathbf{k} \cdot \r\right)},
\end{equation}
\alex{where the symbol $\cdot$ stands for the scalar product between the vectors $\mathbf{k}$ and $\r$.} Each matrix $\R$ can now be projected in the far field at its output via the matrix product
\begin{equation}
\label{Rtt}
    \mathbf{R}_\textrm{out}=\mathbf{T}_0 \times \R,
\end{equation}
Each column of the resulting matrix $\Rout=[R(\kout,\rin)]$ contains the reflected wave-field in the far-field for each input focusing point $\rin$. Injecting Eq.~\ref{RxxMatrix} into the last equation yields the following expression for $\Rout$:
\begin{equation}
\label{Rout}
\Rout = \mathbf{T}_\textrm{out} \times \mathbf{\Gamma} \times \mathbf{H}_\textrm{in},
\end{equation}
where {$\mathbf{T}_\textrm{out}=\mathbf{T}_0 \times \mathbf{H}_\textrm{out}^{\top}$ is the output transmission matrix that describe wave propagation between the focused basis and the Earth surface in the Fourier basis.} In terms of matrix coefficients, the last equation can be rewritten as follows:
\begin{equation}
\label{RkrMatrix}
R_\textrm{out}(\kout,\rin) = \int d\r T_\textrm{out}(\kout,\r) \gamma(\r) {H}_\textrm{in}(\r,\rin).
\end{equation}
\alex{In a multi-target medium made of a few bright scatterers, the reflection matrix can be leveraged to focus selectively on each scatterer. This is the principle of the DORT method [French acronym for Decomposition of the Time Reversal Operator;~\cite{prada,Prada1996}]. Mathematically, the DORT method relies on a singular value decomposition \all{(SVD)} of the reflection matrix. Physically, the singular vectors of $\Rout$ are indeed shown to be the time reversal invariants of the system, i.e the wave-fronts on which would converge an iterative time reversal process, i.e a succession of time reversal operations on the reflected wave-field recorded by the array. }
In the single scattering regime, a one-to-one association \alex{actually} exists between each eigenstate of $\Rout$ and each scatterer. Each singular value is directly equal to the reflectivity $\gamma_i$ of the corresponding scatterer. Each output eigenvector yields the wave-front, { $T_\textrm{out}^*(\kout,\mathbf{r}_i)$,} that should be applied from the far-field in order to selectively focus on each scatterers' position $\mathbf{r}_i$. The DORT method is thus particularly useful for selective focusing in presence of aberrations~\citep{prada,Prada1996} or target detection in a multiple scattering regime~\citep{shahjahan,badon2016,erebus}. However, this decomposition is of poor interest for diffraction-limited imaging since each input eigenvector yields an image of the scatterer,{  $H_\textrm{in}(\mathbf{r}_i,\rin)$,} that is still hampered by aberrations. 

\subsection{Definition and physical interpretation of the distortion matrix}

To cope with the fundamental issue of imaging, a novel operator, the so-called distortion matrix $\mathbf{D}$, has been recently introduced~\citep{Badon2019,william}. Inspired by previous works in ultrasound imaging~\citep{Varslot2004,Robert2008}, it relies on the decomposition of the reflected wavefront into two contributions (see Fig.~\ref{fig3}): (\textit{i}) a geometric component which would be obtained for a point-like target at $\rin$ in a perfectly homogeneous medium (represented by the black dashed line in Fig.~\ref{fig3}b) and which can be directly extracted from the reference matrix $ \mathbf{T}_0$, and (\textit{ii}) a distorted component due to the mismatch between the propagation model and reality (Fig.~\ref{fig3}c). The principle of our approach is to isolate the latter contribution by subtracting, from the reflected wave-front, its ideal counterpart. Mathematically, this operation can be expressed as a Hadamard \all{(element-wise)} product between the normalized reflection matrix $\Rout$ and $\mathbf{T}_0^{*}$,
\begin{equation}
\Dout  =  \Rout \circ  \mathbf{T}_0^{*} ,
\label{D_eq}
\end{equation}   
which, in terms of matrix coefficients, yields{
\begin{equation}
D_\textrm{out}(\kout,\rin)  = {R}_\textrm{out}(\kout,\rin)  T_0^{*}(\kout,\rin)
\label{D_eq_coefficients} .
\end{equation}  } 
The matrix $\Dout=[D(\kout,\rin)]$ connects any input focal point $\rin$ to the distorted component of the reflected wave-field in the far-field. \alex{Removing the ideal phase law predicted by our propagation model from} the reflected wave-field in the Fourier plane as done in Eq.~(\ref{D_eq}) amounts to a change of reference frame. While the original reflection matrix is recorded in the Earth's frame (static underground scanned by the input focusing beam, see Fig.~\ref{fig3}b), the matrix $\Dout$ can be seen a reflection matrix in the frame of the input focusing beam (moving subsoil insonified by a static focusing beam, see Fig.~\ref{fig3}c).

\all{For spatially-invariant aberrations [Eq.~\eqref{iso}], the $\mathbf{D}$-matrix coefficients can be derived from Eqs.~(\ref{RkrMatrix}) and \eqref{D_eq_coefficients}:
\begin{equation}
   D(\mathbf{k}_\textrm{out}, \rin ) = 
  {\tilde{H}}_{\textrm{out}}(\mathbf{k}_\textrm{out}) \int d\mathbf{r}\gamma(\mathbf{r}+\mathbf{r}_\textrm{in}) H_\textrm{in}(\mathbf{r})  e^{i \mathbf{k}_\textrm{out}.\mathbf{r}} ,
  \label{Dalex}
\end{equation}
where $\tilde{H}_\textrm{out}$ is the aberration transmittance, that is to say the 2D Fourier transform of the output PSF $H_\textrm{out}$: ${\tilde{H}}_\textrm{out}(\mathbf{k}_\textrm{out})= \int d\mathbf{r} H_\textrm{out}(\mathbf{r}) e^{-i  \mathbf{k}_\textrm{out}.\mathbf{r}} $.
\all{$\tilde{\mathbf{H}}_{\textrm{out}}$} is the key for optimal focusing since its phase conjugate directly provides the focusing law that needs to be used to overcome the aberrations induced by the medium heterogeneities.}

\all{To extract $\mathbf{\tilde{H}_\textrm{out}}$, the $\mathbf{D}$-matrix is the right tool.  Equation~(\ref{Dalex})} can actually be interpreted as the result of the following fictitious process: Imagine a beam with $\gamma$ as PSF and impinging onto a (fictitious) scatterer at the origin, with reflectivity distribution $H_\textrm{in}(\mathbf{r})$ [see Fig.~\ref{fig3}(c)]; the resulting scattered wavefield in the direction $\kout$ would be given by Eq.~(\ref{Dalex}) \all{[see the analogy with Eq.~(\ref{RkrMatrix})]}. The problem is to isolate \all{the aberration transmittance $\tilde{H}_\textrm{out}$} from the scattering ($\gamma$) and input beam ($H_\textrm{in}$) terms. In that context, iterative time reversal brings a solution : it is well known that it eventually converges the wavefront that overcomes aberrations and optimally focuses on the brightest point of the (fictitious) scatterer~\citep{Robert2008}. In the next section, we show how to implement this idea, to estimate aberration, and finally build a high-resolution image of the subsoil.

\section{Time reversal analysis of the distortion matrix}
~\label{correction}

\all{The subsequent time reversal analysis} consists of different steps that we will describe below. At each iteration, a virtual scatterer is synthesized at input or output through the distortion matrix concept. In the first \alex{step}, a SVD of \alex{$\mathbf{D}_\textrm{out}$} decomposes the \arn{\textit{field-of-view} (\textit{i.e.}, the transverse size of the focal plane)} into a set of isoplanatic patches. The corresponding eigenvectors yield an estimation of the aberration transmittance over each isoplanatic patch. Their phase conjugate provide the focusing laws that enable a (partial) compensation for the phase distortions undergone \all{by the reflected waves during their travel between the focal plane and the geophone array}. {By alternatively applying the same aberration correction process at input, the size of the virtual scatterer can be gradually reduced. \arn{It converges to} optimal focusing laws that will, ultimately, provide a high-resolution mapping of the SJFZ subsoil.}

\subsection{\label{output}Output distortion matrix and isoplanatic patches}
 \begin{figure*}
 \centering
 \includegraphics[width=17cm]{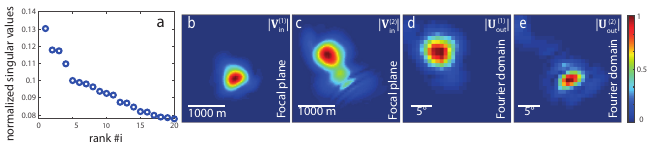}
 \caption{Singular value decomposition of the distortion matrix $\Dout$ at time $\alex{t_B}=4.8$ s and depth $z=$3600 m. (a) Plot of the normalized singular values {$\hat{\sigma}_p$. \all{The two first eigenstates form the relevant signal subspace for imaging. The moduli of the corresponding input eigenvectors, {$\mathbf{V}^{(1)}_\textrm{in}$ (b) and $\mathbf{V}^{(2)}_\textrm{in}$} (c), and output eigenvectors, {$\mathbf{U}^{(1)}_\textrm{out}$ (d) and $\mathbf{U}^{(2)}_\textrm{out}$ (e),} are shown with a linear color scale.} }}
 \label{fig4}
  \end{figure*}

At each depth, a time reversal analysis of the distortion matrix is performed. \all{The first step consists in a \all{SVD} of the output distortion matrix $\mathbf{D}_\textrm{out}$:} 
\begin{equation}
\label{svd}
{\Dout=\mathbf{U}_\textrm{out} \times \mathbf{\Sigma} \times \mathbf{V}_\textrm{in}^\dag}
\end{equation}
or, in terms of matrix coefficients,
\begin{equation}
\label{svd2}
{D(\kout,\rin)=\sum_{i=1}^{N} \sigma_i U_{\textrm{out}}^{(i)}(\kout) V_\textrm{in}^{(i)*}(\rin).}
\end{equation}
{$\mathbf{\Sigma}$ is a diagonal matrix containing the real positive singular values $\sigma_i$ in a decreasing order $\sigma_1>\sigma_2> \cdots >\sigma_N$. $\mathbf{U}_\textrm{out}$ and $\mathbf{V}_\textrm{in}$ are unitary matrices whose columns, $\mathbf{U}_\textrm{out}^{(i)}=[U_\textrm{out}^{(i)}(\mathbf{k_\textrm{out}})]$ and $\mathbf{V}_\textrm{in}^{(i)}=[V_\textrm{in}^{(i)}(\mathbf{r_\textrm{in}})]$}, correspond to the output and input singular vectors, respectively. \all{For spatially-invariant aberrations, 
the physical meaning of this SVD can be intuitively understood by considering the asymptotic case of a point-like input focusing beam [$H_\textrm{in}(x)=\delta(x)$]. In this ideal case, Eq.~(\ref{Dalex}) becomes $ D(\mathbf{k}_\textrm{out}, \mathbf{r}_\textrm{in}  ) =  
{\tilde{H}}_\textrm{out}(u_\textrm{out}) \gamma(\rin)$. 
Comparison with Eq.(\ref{svd2}) shows that 
$\mathbf{D}_\textrm{out}$ is then of rank $1$ -- the 
first output singular vector $\mathbf{U}_\textrm{out}^{(1)}$ yields the aberration transmittance $\mathbf{\tilde{H}}_\textrm{out}$ while the first input eigenvector $\mathbf{V}_\textrm{in}^{(1)}$ directly provides the medium reflectivity. In reality, the input PSF $H_\textrm{in}$ is of course far from being point-like. Moreover, aberration is not laterally invariant across the focal plane. \all{The matrix  $\mathbf{D}_\textrm{out}$ is thus not singular and its spectrum displays a continuum of singular values. Figure \ref{fig4}a confirms this prediction by displaying the normalized singular values, {$\hat{\sigma}_{i}={\sigma}_i/\sqrt{\sum_{j=1}^N  {\sigma}_j^2 }$}, of $\mathbf{D}_\textrm{out}$ at depth $z=3600$ m (${t_B}=4.8$ s). As shown in Supplementary Section S3, only the two first eigenstates are of interest for imaging in this specific case. In the following, we will thus restrict our study to the corresponding signal subspace.}}

\begin{figure}
 \centering
 \includegraphics[width=8.5cm]{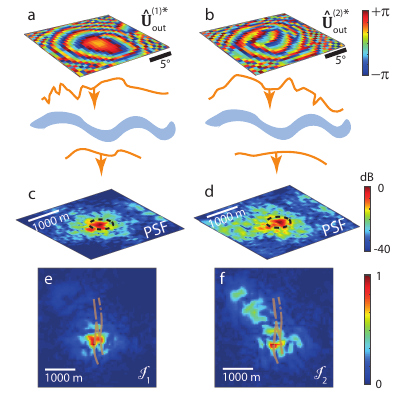}
 \caption{ Output phase distortion correction at time $\alex{t_B}=4.8$ s and depth $z=$3600 m. (a,b) Output focusing laws derived from the phase conjugation of normalized eigenvectors {$\mathbf{\hat{U}}^{(1)}_\textrm{out}$ and $\mathbf{\hat{U}}^{(2)}_\textrm{out}$}, respectively. (c,d) Imaging PSF extracted from the main antidiagonals of the focused reflection matrices \all{$\mathbf{R}_1$ and $\mathbf{R}_2$}, respectively (Eq.~\ref{Rp}). The color scale is in dB. (e,f) \all{Corresponding confocal images $\mathcal{I}_1$ and $\mathcal{I}_2$}, respectively.}
 \label{fig5}
  \end{figure}

\all{In a medium displaying a complex wave velocity distribution, the SVD of $\mathbf{D}_\textrm{out}$ can provide a decomposition of the field-of-view into several isoplanatic patches. On the one hand, each input singular vector, {$\mathbf{V}^{(p)}_\textrm{in}$}, maps onto the corresponding isoplanatic patch $p$. Figures~\ref{fig4}b and c confirm this assertion by showing that {$\mathbf{V}^{(1)}_\textrm{in}$} and $\mathbf{V}^{(2)}_\textrm{in}$ focus onto two disjoint areas at depth $z=3600$ m. On the other hand, each output singular vector {$\mathbf{U}^{(p)}_\textrm{out}$} is linked to the corresponding aberration transmittance $\tilde{H}^{(p)}_\textrm{out}$~\citep{Badon2019}:{
\begin{equation}
\label{U1}
 {U}^{(p)}_\textrm{out}(\kout) \propto \tilde{H}^{(p)}_\textrm{out}(\kout) \left [ \tilde{H}^{(p)}_\textrm{in} \circledast \tilde{H}^{(p)}_\textrm{in} \right ](\kout),
\end{equation}} }  
\all{where the symbol $\circledast$ stands for the correlation product. {However, $\mathbf{U}^{(p)}_\textrm{out}$} is also modulated by the autocorrelation function $\left [ \tilde{H}^{(p)}_\textrm{in} \circledast\tilde{H}^{(p)}_\textrm{in} \right] $ (see Fig~\ref{fig3}d). This last term is a manifestation of the finite size $\delta_\textrm{in}^{(p)}$ of the virtual reflector (Fig.~\ref{fig3}d) that tends to limit the support of the eigenvector {$\mathbf {U}^{(p)}_\textrm{out}$} to \arn{$\Delta k^{(p)}_\textrm{out} \sim \lambda z/\delta_\textrm{in}^{(p)}$}.} Figures~\ref{fig4}d and e confirm this theoretical prediction by showing the modulus of {$\mathbf {U}^{(1)}_\textrm{out}$} and {$\mathbf {U}^{(2)}_\textrm{out}$}, respectively. Both {singular vectors} cover a restricted and different angular domain in the Fourier space \arn{($\Delta \theta_\textrm{out}^{(p)} = \Delta k_\textrm{out}^{(p)}/k_0 \sim 10^\textrm{o}$)}. To circumvent this issue, one trick is to use only the phase of these eigenvectors {$\mathbf {U}^{(p)}_\textrm{out}$} (Figs.~\ref{fig5}a and b) by considering the normalized vector {$\mathbf {\hat{U}}^{(p)}_\textrm{out}$, such that 
\begin{equation}
\label{normU}
{\hat{U}}^{(p)}_\textrm{out}(\kout)=U^{(p)}_\textrm{out}(\kout)/|U^{(p)}_\textrm{out}(\kout)|.
\end{equation}}
Indeed, if we make the realistic hypothesis of a real and positive autocorrelation function {$\left [\tilde{H}^{(p)}_\textrm{in} \circledast \tilde{H}^{(p)}_\textrm{in}\right]$} in Eq.~(\ref{U1}), the normalized vector {$\mathbf{\hat{U}}^{(p)}_\textrm{out}$} should, in principle, provide the aberration transmittance {$\mathbf{\tilde{H}}^{(p)}_\textrm{out}$}. In practice, noise degrades the estimation of {$\mathbf{\tilde{H}}^{(p)}_\textrm{out}$} outside the coherence area \arn{$\Delta k_\textrm{out}$}. In the following, we will note { $\delta\mathbf{\tilde{H}}^{(p)}_\textrm{out}=\mathbf{{\hat{U}}}^{(p)*}_\textrm{out} \circ \mathbf{\tilde{H}}^{(p)}_\textrm{out}$}, the residual phase mismatch between{ $\mathbf{{\hat{U}}}^{(p)}_\textrm{out}$ and $\mathbf{\tilde{H}}^{(p)}_\textrm{out}$}. 

Despite this phase mismatch, the phase conjugate of {$\mathbf{\hat{U}}^{(p)}_\textrm{out}$} can be used as a focusing law to compensate (at least partially) for phase distortions at the output (Fig.~\ref{fig3}f). \all{Updated focused reflection matrices can indeed be obtained as follows:
\begin{equation}
\label{Rp}
    \mathbf{R}_p= (\mathbf{\hat{U}}^{(p)}_\textrm{out} \circ \mathbf{T_0})^{\dag}   \times \Rout
\end{equation}
New confocal images, $\mathcal{I}_p(\mathbf{r},z)=|R_p(\mathbf{r},\mathbf{r},z)|^2$, can be extracted from the diagonal of {$\mathbf{R}_p$} after output aberration compensation.} The result is displayed in Figs.~\ref{fig5}e and f at depth $z=3600$ m. The comparison with the initial confocal image (Fig.~\ref{fig2}c) illustrates the benefit of our matrix approach. While the original image displays a random speckle feature across the field-of-view, \all{$\mathcal{I}_1$} reveals a complex structure in the vicinity of the surface traces of the Clark Fault. On the contrary, the image \all{$\mathcal{I}_1$} spans along an oblique direction compared to the direction of the fault. While it shows a more diffuse image of the Clark Fault at the center of the field-of-view, \all{$\mathcal{I}_2$} reveals a strong scattering structure on the west of the fault. 

To quantify the gain in image quality, the local imaging PSF should be investigated. To that aim, the antidiagonals of $\mathbf{R}_p$ can be used to probe the local imaging PSF across the field-of-view [Eq.~(\ref{IrDr_spec})]. The resulting PSFs \arn{are} displayed in Figs.~\ref{fig5}c and d. It should be compared with the initial imaging PSF (Fig.~\ref{fig2}b). While the original PSF exhibits a distorted central lobe that spans over almost four diffraction-limited \alex{transverse} resolution cells (white circle in Fig.~\ref{fig2}b), the corrected PSF shows a central lobe thinner than the diffraction limit. This striking result will be discussed further. Note, however, the occurence of a strong secondary lobe and an incoherent background that can be accounted for by the subsistence of input aberrations. The latter ones are tackled in the next section.

\subsection{Input distortion matrix and super-resolution}
 \begin{figure}
 \centering
 \includegraphics[width=8.5cm]{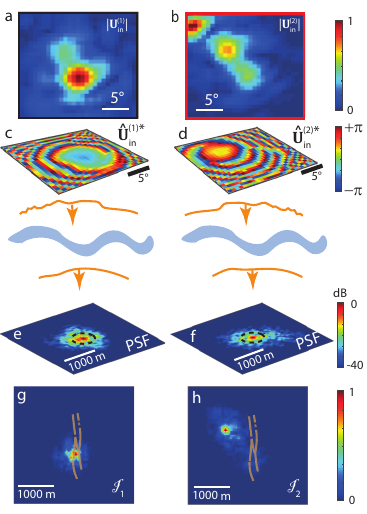}
 \caption{ Input phase distortion correction at time $\alex{t_B}=4.8$ s and depth $z=$3600 m. (a,b) Modulus of {input} eigenvectors {$\mathbf{U}_\textrm{in}^{(1)}$ and $\mathbf{U}_\textrm{in}^{(2)}$.} (c,d) Output focusing laws derived from the phase conjugation of normalized eigenvectors {$\mathbf{\hat{U}}_\textrm{in}^{(1)}$ and $\mathbf{\hat{U}}_\textrm{in}^{(2)}$}, respectively. (e,f) Imaging PSF extracted from the main antidiagonals of the focused reflection matrices \all{$\mathbf{R}_1$ and $\mathbf{R}_2$}, respectively [Eq.~(\ref{Rp2})]. The color scale is in dB. (g,h) Corresponding confocal images  \all{$\mathcal{I}_1$ and $\mathcal{I}_2$ }, respectively.}
 \label{fig6}
  \end{figure}
\all{Aberrations undergone by the incident waves are now compensated by building input distortion matrices {$\mathbf{D}^{(p)}_\textrm{in}$} (Fig.~\ref{fig3}c):
\begin{equation}
\label{Din}
\mathbf{D}_\textrm{in}^{(p)}= \mathbf{T}_0^{\dag} \circ \left (  \mathbf{R}_p \times \mathbf{T_0^\top} \right ) .
\end{equation}}
{The SVD of each matrix $\mathbf{D}_\textrm{in}^{(p)}$ is performed. Among all the output eigenvectors of these matrices, one is of special interest: their $p^{th}$ eigenvector $\mathbf{U}_\textrm{in}^{(p)}$. Its modulus is shown for $p=1$ and $2$} in Fig.~\ref{fig6}a and b, respectively. {The input eigenvectors $\mathbf{U}_\textrm{in}^{(p)}$ exhibit a much larger angular aperture ($\Delta \theta_\textrm{in} \sim 20^\textrm{o}$) than the output eigenvectors  $\mathbf{U}_\textrm{out}^{(p)}$} ($\Delta \theta_\textrm{out}\sim 10^\textrm{o}$). \all{As $\mathbf{U}_\textrm{out}^{(p)}$ [Eq.~(\ref{U1})], the input singular vectors {$\mathbf{U}_\textrm{in}^{(p)}$} can indeed be expressed as follows:}{
\begin{equation}
    \label{U1in}
{U}_\textrm{in}^{(p)} (\kin) \propto \tilde{H}^{(p)}_\textrm{in}(\kin)  \tilde{H}^{(p)}_\textrm{in}(\kin) \left [ \delta\tilde{H}^{(p)}_\textrm{out} \circledast \delta\tilde{H}^{(p)}_\textrm{out} \right ](\kin).
\end{equation}}
 As mentioned before, the correlation term{, $\delta\hat{H}^{(p)}_\textrm{out} \circledast \delta\hat{H}^{(p)}_\textrm{out}$,} can be seen as the Fourier transform of the virtual scatterer synthesized from the output focal spots in the distortion matrix {$\mathbf{D}^{(p)}_\textrm{in}$} (see Fig.~\ref{fig3}c). The width {$\delta^{(p)}_\textrm{out}$} of the corrected output PSF {$\delta H^{(p)}_\textrm{out}$} (Figs.~\ref{fig5}c and d) being much thinner than the original one $\delta^{(0)}_\textrm{in}$ at the input (Fig.~\ref{fig2}b), the correlation width {$\Delta k_\textrm{in}^{(p)}\sim \lambda z/\delta^{(p)}_\textrm{out}$} of the incident wave-field is larger than the output correlation width {$\Delta k^{(p)}_\textrm{out}\sim \lambda z/\delta^{(0)}_\textrm{in}$} of the reflected wave-field at the previous step. 

The phase of the first input singular vectors {$\mathbf{U}_\textrm{in}^{(p)}$} (Figs.~\ref{fig6}c and d) are thus better estimators of {$\mathbf{\tilde{H}}_\textrm{in}^{(p)}$} than the original input eigenvectors {$\mathbf{U}^{(p)}_\textrm{out}$} \all{for {$\mathbf{\tilde{H}}_\textrm{out}^{(p)}$}}. The phase mismatch between them will be noted {$\delta\mathbf{\tilde{H}}_\textrm{in}^{(p)}$} in the following. Despite this residual phase error, the normalized eigenvectors, {$\mathbf{\hat{U}}_\textrm{in}^{(p)}$}, can be used as focusing laws to compensate for input aberrations. \all{The resulting focused reflection matrices,
\begin{equation}
\label{Rp2}
    \mathbf{R}_p=   \left [  \mathbf{T}_0^\top \circ   \mathbf{U}^{(p)*}_\textrm{in} \circ  \Din^{(p)}  \right ] \times \mathbf{T}_0^{*} 
\end{equation}
yields novel confocal images $\mathcal{I}_p$ of the subsoil reflectivity. The result is} displayed in Figs.~\ref{fig6}g and h at depth $z=3600$ m. Their comparison with the previous images (Figs.~\ref{fig5}e and f, respectively) illustrates the benefit of a simultaneous aberration correction at input and output. While \all{$\mathcal{I}_1$} yields a refined image of the complex structure lying in the first isoplanatic patch (Fig.~\ref{fig6}g), the second eigenvector \all{$\mathcal{I}_2$} now clearly highlights a coherent reflector belonging to a second isoplanatic patch on the west of the fault (Fig.~\ref{fig6}h).  

The \alex{transverse} resolution and contrast enhancements are now quantified from the antidiagonals of the updated reflection matrices {$\mathbf{R}_p$} after input and output aberration compensation. An imaging PSF can be deduced for each isoplanatic patch $p$ by considering the antidiagonal of {$\mathbf{R}_p$} whose common mid-point {corresponds to} the maximum of each image{ $\mathcal{I}_p$} [Eq.~(\ref{IrDr_spec})]. The result is displayed in Figs.~\ref{fig6}g and h. The strong secondary lobes exhibited by the imaging PSF at the previous step (Figs.~\ref{fig5}c and d) have been fully suppressed thanks to the compensation of aberrations at the input. Strikingly, the spatial extension {$\delta_\textrm{in}^{(p)}$} of the imaging PSF at -6dB is of the order of 150 m. This value is much thinner than the expected \alex{lateral} resolution cell {$\delta_0 \sim  600$ m}, depicted as a white dashed line in Figs.~\ref{fig6}e and f. The observed \alex{transverse} resolution is one-fourth of the theoretical limit based on the geophone array aperture. 

Two reasons can be invoked to account for that surprising result. The first possibility is an overestimation of the wave speed $c$ \mich{at shallow depth. This value actually dictates the maximum spatial frequency exhibited by body waves induced at the Earth surface.} {However, the chosen value $c_0=1500$ m/s seems in agreement with P-wave velocity measured in the shallow ($<$100 m) fault zone at the site under study~\citep{share2020}.} The other explanation is that the heterogeneous subsoil acts as a lens between the geophones and the focal plane either by scattering or wave guiding. {As in a time reversal experiment, scattering by small-scale heterogeneities can widen the angular aperture of the focused beam~\citep{derode0}.
\all{~\cite{khaidukov2004} also discussed that the scattering component of the wavefield holds information on small-scale subsurface heterogeneities and therefore contributes to the high resolution or even superresolution of the migrated images.}
Alternatively, the local gradient of the background seismic velocity in the vicinity of the fault can constitute a wave guide through which seismic waves can be channeled\mich{~\citep{Li1990}. Similarly to scattering, reflections on wave guide boundaries can also enlarge the effective array aperture: This is the so-called kaleidoscopic effect~\citep{Roux2001}.} In both cases,} if the induced aberrations are properly compensated, the spatial extension $\delta_{in/out}^{(p)}$ of the imaging PSF can be improved compared to the diffraction-limited \alex{lateral} resolution $\delta_0$ predicted for a homogeneous medium. These physical phenomena and their impact on imaging will be discussed further in Sec.~\ref{sec:discussion}.


\subsection{Normalized correlation matrix and compensation for high-order aberrations}
 \begin{figure}
 \centering
 \includegraphics[width=8.5cm]{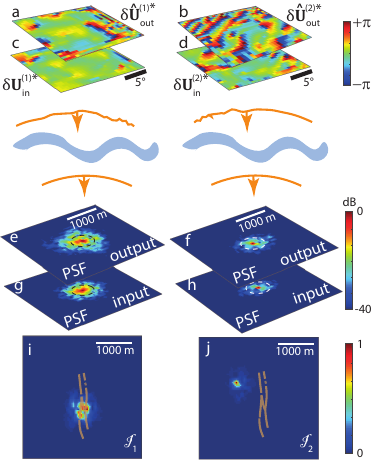}
 \caption{ Correction of residual phase distortions by making the virtual scatterer point-like \alex{($t_B=4.8$ s, $z=3600$ m)}. (a,b,c,d) Output and input focusing laws derived from the phase conjugation of the normalized eigenvectors {$\delta\mathbf{\hat{U}}_\textrm{out}^{(1)}$, $\delta\mathbf{\hat{U}}_\textrm{out}^{(2)}$, $\delta\mathbf{\hat{U}}_\textrm{in}^{(1)}$ and $\delta\mathbf{\hat{U}}_\textrm{in}^{(2)}$}, respectively. (e,f,g,h) Imaging PSFs extracted from the main antidiagonal of the focused reflection matrices {$\mathbf{R}_p$} after application of the additional focusing law displayed in (a,b,c,d), respectively. The color scale is in dB. (i,j) Final confocal images built from the diagonal of the final reflection matrices $\mathbf{R}_p$, respectively [Eq.~(\ref{Rpf})].}
 \label{fig7}
  \end{figure}
Despite these remarkable properties, the imaging PSFs in Figs.~\ref{fig6}e and f still exhibit high-order aberrations that result in an incoherent background of -20 dB beyond the \alex{transverse} resolution cell. To compensate for these residual aberrations and again improve the image quality, a last step consists in considering the normalized correlation matrix \all{of the residual wave-front distortions} [see \all{Supplementary Section S4}]. 

\all{It first consists in building, from the updated reflection matrix \all{$\mathbf{R}_p$} (Eq.~\ref{Rp2}), a novel output distortion matrix $\delta \mathbf{D}^{(p)}_\textrm{out}$ :
{
\begin{equation}
\delta \mathbf{D}_\textrm{out}^{(p)}= \left [ \mathbf{T_0}   \times \mathbf{R}_p \right] \circ  \mathbf{T}_0^{*} ,
\end{equation}}
The corresponding correlation matrix, {$\delta \mathbf{C}_\textrm{out}^{(p)}=\delta\mathbf{D}^{(p)}_\textrm{out} \times \delta\mathbf{D}_\textrm{out}^{(p)\dag}$}, is then computed and its coefficients are normalized, such that:} 
\begin{equation}
\label{hat}
   \delta \hat{C}_\textrm{out}^{(p)}(\kout,\kpout)={\delta{{C}}_\textrm{out}^{(p)}(\kout,\kpout)}/{{|{\delta C_\textrm{out}^{(p)}}}(\kout,\kpout)|}.
\end{equation}
As illustrated by Fig.~\ref{fig3}e, this operation makes the virtual reflector point-like. Such a matrix is \all{actually} equivalent to the time reversal operator associated with a point-like reflector at the origin~[\cite{william}, see \all{Supplementary Section S4}].,  In that case, the matrix {$\delta \mathbf{\hat{C}}^{(p)}_\textrm{out}$} is \all{ideally} of rank 1 and its eigenvector {$\delta \mathbf{U}^{(p)}_\textrm{out}$} yields the residual aberration phase transmittance:{
\begin{equation}
\delta \mathbf{U}^{(p)}_\textrm{out} (\kout)= \delta \tilde{H}^{(p)}_\textrm{out}(\kout)
\end{equation}}
The phase of the eigenvectors {$\delta \mathbf{U}^{(p)}_\textrm{out}$} are displayed in Fig.~\ref{fig7}a and b, respectively. Compared to the Fresnel zone fringes exhibited by the first-order aberrations in {$\mathbf{U}^{(p)}_\textrm{out}$} (Fig.~\ref{fig5}a and b), {$\delta \mathbf{U}^{(p)}_\textrm{out}$} displays higher-order aberrations, thereby leading to more complex focusing law. \all{Updated focused reflection matrices {$\mathbf{R}_p$ are then deduced:
\begin{equation}
\label{Rpf}
\mathbf{R}_p=
\mathbf{T}_0^{\dag} \times \left [  \mathbf{T}_0 \circ \delta\mathbf{ U}^{(p)*}_\textrm{out} \circ \delta \Dout^{(p)}  \right ] .
\end{equation}}}
\all{Their main antidiagonal enables an estimation of the imaging PSF over each isoplanatic patch (see Figs.~\ref{fig7}e and f).} This second correction step drastically reduces the spatial extension of the PSF width compared to the previous step (Figs.~\ref{fig6}e and f). The incoherent background is below -30 dB beyond the theoretical \alex{transverse} resolution cell ($\delta_0$= 600 m at depth $z=3600$ m){. Moreover, the FWHM of the PSF is now of the order of the wavelength: $\delta_\textrm{out}^{(p)} \sim \lambda \sim 100$ m. Strikingly,} the observed \alex{lateral} resolution is one-sixth of the theoretical limit based on the geophone array aperture.

As before, the process can be iterated by exchanging the focused and Fourier basis at input and output. A novel input distortion matrix {$\delta \mathbf{D}_\textrm{in}^{(p)}$} is built for each isoplanatic patch $p$. {Additional} input focusing laws {$\delta\mathbf{U}_\textrm{in}^{(p)*}$} are extracted through the \all{SVD} of the normalized correlation matrix {$\delta \mathbf{\hat{C}}_\textrm{in}^{(p)}$}. The \all{corresponding wave-fronts} are displayed in Figs.~\ref{fig7}c and d. \all{They exhibit phase fluctuations that are reduced compared to} their output counterpart \all{(Figs.~\ref{fig7}a and b)}, \arn{an indication} that our approach gradually converges towards a finite aberration phase law. The normalized eigenvectors, {$\delta\mathbf{\hat{U}}_\textrm{in}^{(p)}$}, can be used as input focusing laws that should be applied to {$\mathbf{D}_\textrm{in}^{(p)}$} to compensate for residual input aberrations. The resulting images of the subsoil reflectivity are displayed in Figs.~\ref{fig7} i and j at $z=3600$ m over the two main isoplanatic patches. Their comparison with the previous images (Figs.~\ref{fig6}g and h, respectively) illustrates the benefit of iterating the aberration correction process. The first isoplanatic patch highlights the presence of three distinct reflectors that arise in the vicinity of the fault's surface traces (Fig.~\ref{fig7}i). The second isoplanatic patch yields a highly-resolved image of a coherent reflector on the west of the Clark fault (Fig.~\ref{fig7}i). Compared to the previous image in Fig.~\ref{fig7}j, the contrast is drastically improved. This observation can be understood by looking at the corresponding PSFs (Figs.~\ref{fig7}g and h). The incoherent background is below -35 dB beyond the theoretical \alex{transverse} resolution cell ($\delta_0$= 600 m at depth $z=3600$ m). {Last but not least,} the FWHM of the PSF nearly reaches the diffraction limit $\delta^{(p)}_\textrm{in} \sim 80$ m. {Remarkably,} the observed \alex{lateral} resolution is nearly one-eighth of the theoretical limit based on the geophone array aperture. {As mentioned in the previous section, the reasons for this spectacular resolution will be discussed in Sec.~\ref{sec:discussion}.}
  
The comparison of these final images with the original one (Fig.~\ref{fig2}c) illustrates the relevance of {a} matrix approach for deep seismic imaging. In the next section, the three-dimensional structure of the SJFZ is revealed by \arn{combining} the images derived at each depth. 
At last, based on the derived high-resolution 3D images, a structural interpretation of the SJFZ is provided. 

\section{Three-dimensional imaging of the San Jacinto Fault zone}
 	 \begin{figure}
 \centering
 \includegraphics[width=8cm]{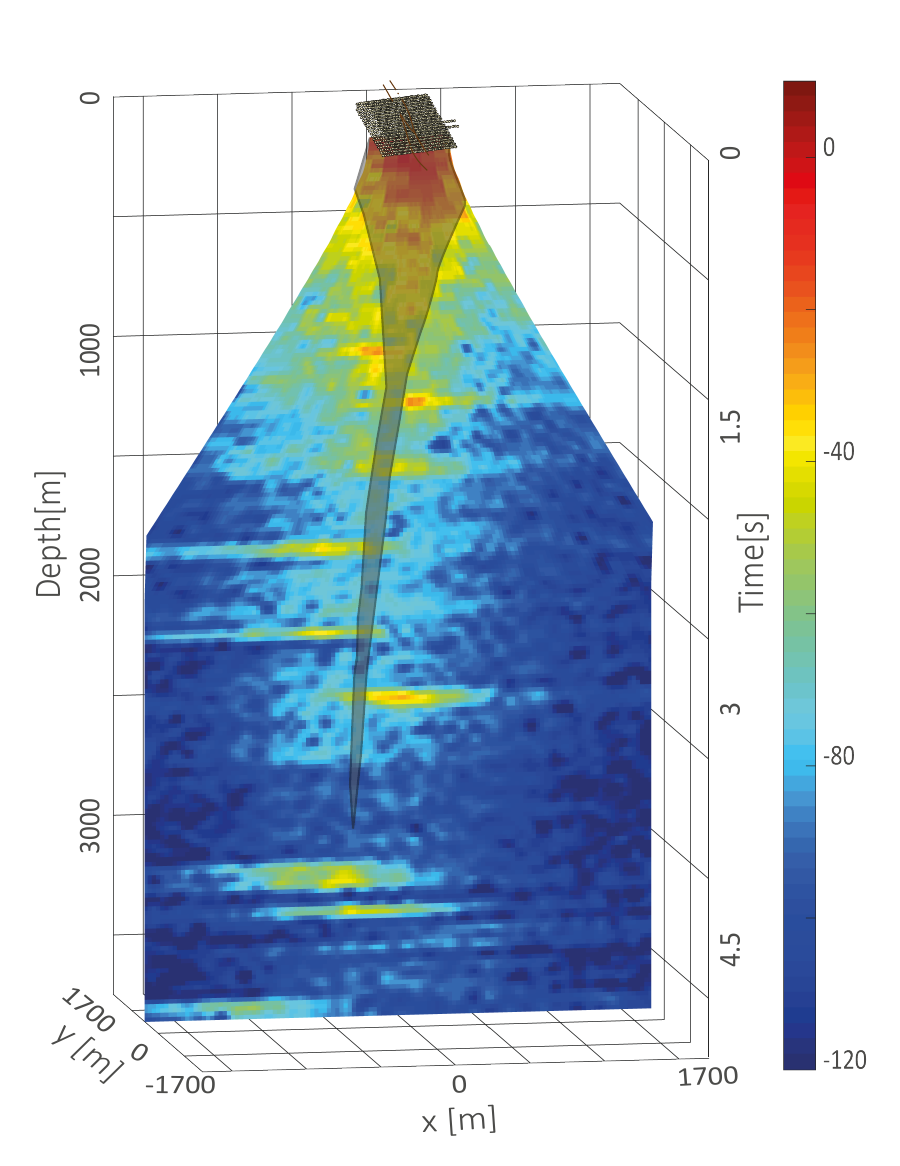}
 \caption{Vertical slice of the 3D confocal image obtained from stacked corrected images derived at each depth. The slice orientation is chosen to be normal to the fault plane. \alex{From this image, the fault location can be circumscribed and is represented by the shaded area.} The color scale is in dB.}
 \label{fig8}
  \end{figure}
  
Having demonstrated in Sec.~\ref{correction} how to correct phase distortions {at each depth,}
a 3D image of the subsurface can now be uncovered. To that aim, each isoplanatic patch should be recombined at each depth. The resulting image $\mathcal{I}_M$ is a coherent sum between the diagonal coefficients of the corrected reflection matrices $\mathbf{R}_p$:
\begin{equation}
\label{immat}
\mathcal{I}_M\left(\mathbf{r},z\right)\equiv \left|\sum_{p=1}^P \all{R_{p}}\left(\r,\r,z\right)\right|^2.
\end{equation}
\al{In the present case, the number $P$ of isoplanatic patches is found to be equal to 2 over the whole depth range. The fault structure, showing a different wave velocity distribution on each side of the fault, probably explains this peculiar behavior.
}

{Figure~\ref{fig8} shows} 
a slice of the final 3D image \alex{$\mathcal{I}_M\left(\mathbf{r},z\right)$} with the same orientation as the original \alex{cross-sectional view} displayed in Fig.~\ref{fig2}\alex{d}. 
While the raw image $\mathcal{I}$ is completely blurred, the gain in \alex{horizontal} resolution provided by the matrix aberration correction process reveals {an image of the SJFZ subsoil with} a refined level of details. Based on the surface traces {displayed in} Fig. \ref{fig1}a, the fault is expected to 
{spread} orthogonally to the $x$-axis. The choice of the profiles' orientation in Figs.~\ref{fig2}\alex{d} and \ref{fig8} {is dictated by our willingness to 
highlight the fault blocks on the right and left side of the slip interface. }

{The SJFZ is} a major continental strike-slip fault system {that} accumulated through history a significant slip of \arn{tens} of kilometers. Such large fault systems induce zones of strongly damaged materials~\citep[and references therein]{BZ2003} and the damage is expected to be pronounced in the shallow crust. {Thanks to the distortion matrix correction,} Fig.~\ref{fig8} shows the two main signatures of the fault in diffraction imaging. First, it reveals a damage zone associated with a dense distribution of scatterers. 
The damage volume extends from the surface to \alex{1000} m below with a 
{section decreasing in depth}. {Second, beyond a depth of \alex{1000} m, the matrix image displays strong echoes associated with the presence of sub-horizontal reflectors}. The corresponding {strata} layers are located at different depths on both sides of the fault. This discontinuity {seems to indicate the fault location in depth \alex{(see the shaded area in Fig.~\ref{fig8})}, although the damage area is no longer discernible beyond \alex{1000} m}. This example thus shows both the efficiency of {our approach to cope with phase distortions} and the potential of passive {seismic} imaging to study the fine structure of active faults {in depth}.

{In this section, only a slice of the 3D volume has been shown to illustrate the drastic improvements granted by a matrix approach of seismic imaging (Fig.~\ref{fig8}). Its comparison with the raw confocal image highlights the benefit of a drastic gain in \alex{horizontal} resolution (Fig.~\ref{fig2}c). A more detailed interpretation of the subsurface properties will be the focus of a future study. Indeed, a detailed interpretation of the obtained 3D image and a confrontation with previous studies may provide information on the structural properties of the fault zone. More generally, the high \alex{transverse} resolution and penetration depth of our matrix imaging method will play an important role in detecting active faults, evaluating their long-term behaviour and, consequently, following the deformation of the Earth's crust. It will also be an essential tool for understanding earthquake physics and evaluating seismic hazard.}

\section{Discussion}
\label{sec:discussion}
\alex{Although the image of the SJFZ fault displayed by Fig.~\ref{fig8} is \al{encouraging}, our matrix imaging approach still suffers from some limitations in its current form. First, it should be noted that the available velocity model is a \alex{very} rough approximation of reality. A constant velocity of 1500 m/s was chosen since it yields the best image (i.e optimized resolution) over a larger depth range. Here, the optimal velocity model is probably lower in the damage zone and gradually increasing beyond. Figure~S4 in the Supplementary Material confirms that intuition by showing the final images for three different wave velocity models. While the image obtained for $c_0=1000$ m/s seems to provide a better axial resolution in the damage zone ($t_B<1$ s), the image built from $c_0=2000$ m/s seems to be much better at large depths ($t_B>3.5$ s).}

\alex{The time gating operation performed in Eq.~\ref{Rrr_broadband} means that the echoes considered at each depth are associated with scattering events that exhibits a time of flight contained in the window $[t_B-\delta t/2;t_B+\delta t/2]$. In a layered medium, the corresponding isochronous volume is an horizontal layer of thickness  $\delta z \sim c\delta t$ centered around a coherence plane located at $\bar{z} \sim \bar{c} t_B/2$, with $\bar{c}$ the integrated speed-of-sound such that $\bar{c}^{-1}=z^{-1}\int_0^z [c(z)]^{-1} dz$. The depth $z=c_0 t_B/2$ imposed by the propagation model is thus likely to be inaccurate. Yet a correct model \alex{would improve the vertical resolution and} dilate the subsboil 3D image up or down but would not change significantly its transverse evolution. }

\alex{For a more complex distribution of seismic wave velocity (such as an anticline), aberrations will also distort the coherence surface and the associated isochronous volume. The distortion matrix method will then correct aberrations along this coherence surface but the vertical (axial) aberrations and the distortion of the coherence surface will remain unchanged. To cope with this issue, a first option is to consider the impulse responses between virtual geophones located at different depths. Such a 3D reflection matrix approach will be addressed in future works.  A second strategy is to include the matrix approach in any inversion scheme that aims at mapping the velocity distribution of bulk seismic waves in the underground. This would, in turn, correct the image from its axial aberrations. Indeed, a wave velocity model closer to reality and a more accurate approximation of the propagating Green's function (Eq.~\ref{Gfreespace}) will always improve the final image, especially its vertical resolution.}

\alex{Second, our approach relies on a scalar model of wave propagation. In the present case, this approximation is justified by the fact that we are only considering the vertical components of the computed \al{impulse responses} between geophones. As the bulk waves are mostly propagating in the vertical direction, the detected echoes are mainly associated with P waves. Nevertheless, possible shear wave conversion during wave propagation can occur. Moreover, a more contrasted image could be obtained if we were able to consider the other components of the Green's functions and also take into account the presence of shear waves in our propagation model. The method could thus be refined in the near future by taking into account both P- and S-waves as well as potential wave conversion between them induced by scattering. However, these aspects are out-of-scope for a first demonstration in the context of the seismic imaging of a fault zone.}

\alex{Despite the limits of the propagation model used in our matrix approach, the 3D image of the SJFZ subsoil exhibits striking properties that are subject to physical interpretation. In a layered medium,  constant horizontal slowness implies that the structure outside of the lateral extension of the receiver area cannot be imaged. On the contrary, in Fig.~\ref{fig8}, the horizontal strata layers are imaged on each side of the fault over a field-of-view much larger than the geophone array dimension. Two reasons can account for this surprising result. First, the scattering between each layer may not be only specular but also induced by a distribution of localized inhomogeneities at each layer interface. The transverse images shown at depth z=3600 m in Fig.~\ref{fig7} confirm this by highlighting the presence of four localized scattering structures in the first isoplanatic patch [Fig.~\ref{fig7}(i)]. Second, the large extension of the image in Fig.~\ref{fig8} can also be due to the scattering induced by the strongly heterogeneous damage area. The effective geophone array aperture can thus be increased by scattering and so is the imaged area. }

\alex{Last but not least, the matrix image of the SJFZ fault shows a transverse super resolution highlighted by the imaging PSF displayed in Fig.~\ref{fig7}(e)-(h). A first reason for this striking result could be an overestimation of the wave speed $c$ \mich{at shallow depth. This value actually dictates the maximum spatial frequency exhibited by body waves induced at the Earth surface.} However, the chosen value $c_0=1500$ m/s seems in agreement with P-wave velocity measured in the shallow ($<$100 m) fault zone at the site under study~\citep{share2020}. Alternatively, the local gradient of the background seismic velocity in the vicinity of the fault can constitute a wave-guide through which seismic waves can be channeled~\citep{Li1990}. Reflections on wave guide boundaries can also enlarge the effective array aperture: This is the so-called kaleidoscopic effect~\citep{Roux2001}. As mentioned above, a last hypothesis is that the heterogeneous subsoil acts as a lens between the geophones and the focal plane either by scattering or wave guiding. As in a time reversal experiment, scattering by small-scale heterogeneities can widen the angular aperture of the focused beam~\citep{derode0}. Here the damage area is particularly heterogeneous and can play the role of scattering lens. Moreover, its location near the surface and its finite thickness ($\sim 1000$ m) implies the existence of an angular memory effect even for multiple scattering speckle~\citep{freund,feng,katz_2012,katz_2014}. In that configuration, multiple scattering manifests itself as high-order aberrations associated with relatively small isoplanatic patches. Such high-order aberrations can be corrected by our matrix method in the plane wave basis~\citep{Badon2019}. This justifies \textit{a posteriori} the choice of this basis for the correction of aberrations. Moreover, plane wave beamforming is particularly adequate in a multi-layered medium since aberrations are laterally-invariant in that frame. Beyond the specific case of SJFZ, note that the choice of basis for the aberration correction is flexible\all{~\citep{lambert_ieee2}}. The reflection matrix can be ideally projected onto any aberrating layer in the subsoil. This choice shall be  dictated by the local topography and any prior knowledge on the local distribution of seismic wave velocities in the zone under study.}

\al{Multiple scattering and/or wave-guiding effects could also explain the optimal velocity ($c_0=1500$ m/s) found for the wave propagation model. Initially, this choice was justified by the focusing quality and image resolution reached at the end of the matrix imaging process (see Supplementary Figs.~S3 and S4). Nevertheless, a physical interpretation can now be provided to account for it. Our hypothesis is that a weak wave velocity model can enable the time gating of multiply-scattered or guided waves that are, ultimately, transmitted until the focal plane where they can be efficiently scattered by subsoil heterogeneities. Indeed, such distorted paths are associated with larger echo times than the ballistic waves usually considered by reflection imaging methods. As mentioned above, they can enlarge the angular aperture of our imaging system and account for a transverse resolution much better than what would be expected if we had used direct ballistic waves. In other words, the use of a relatively weak wave velocity allows us to adapt a rough homogeneous model to a medium that displays a strongly heterogeneous wave velocity distribution. This remains of course an hypothesis and a further analysis will be required in order to prove rigorously the origin for the transverse super resolution in the context of fault seismic imaging.}

\section{Conclusion}
{Inspired by pioneering works in optical microsopy~\citep{badon2016,Badon2019} and ultrasound imaging~\citep{Robert2008,william}, a novel matrix approach to seismic imaging is proposed in this paper. Taking advantage of the reflection of bulk seismic waves by heterogeneities in depth, it can be applied to both active or passive seismic imaging. The strength of this approach lies in the fact that it works even when the velocity distribution of the subsoil is unknown. By projecting the seismic data either in the focused  basis or in the Fourier plane, it takes full advantage of all the information contained in the collected data. For aberration correction, projection of the reflection matrix into a dual basis allows the isolation of the distorted component of the reflected wave-field. Seen from the focused basis, building the distortion matrix $\mathbf{D}$ consists in virtually shifting all the input or output focal spots at the onto the same virtual location. An iterative time reversal analysis then allows to unscramble the phase distortions undergone by the wave-front during its travel between the Earth surface and the focal plane. A one-to-one association is actually found between each eigenstate of $\mathbf{D}$ and each isoplanatic patch in the field-of-view. More precisely, each singular vector in the Fourier space yields the far-field focusing law required to focus onto any point of the corresponding isoplanatic patch. A confocal image of the subsoil reflectivity can then be retrieved as if the underground had been made homogeneous. }

In this paper, as a proof-of-concept, the case of the San Jacinto Fault zone is considered. While a raw confocal image suffers from an extremely bad \alex{horizontal} resolution due to the strong lateral variations of the seismic velocities in the vicinity of the fault, our matrix approach provides a high-resolution image. Strikingly, its \alex{lateral} resolution is almost one eighth below the diffraction limit imposed by the geophone array aperture. \alex{This surprising property may be accounted for by the heterogeneities of the subsoil that can play the role of a scattering and/or channeling lens which increases drastically the effective array aperture.} The matrix image reveals a damage volume particularly pronounced in the shallow crust \alex{($<$1000 m)}. At larger depth, the 3D image of the fault exhibits the fault blocks on the right and left side of the slip interface.  A more detailed interpretation of the obtained image will be the focus of a future study. 

Besides seismic fault zones, a matrix approach of passive imaging is particularly suited to the study of volcanoes. While the multiple scattering problem in volcanoes has been recently tackled under the reflection matrix approach~\citep{erebus}, the distortion matrix would be particularly powerful to restore a diffraction-limited image of the in-depth structure of a volcano. Additionally, our matrix method \all{could} advantageously be applied to real-time geophysical imaging, similarly to adaptive optics in astronomy. This may provide a valuable tool for monitoring oil or gas reservoirs when drilling, extracting hydrocarbon or injecting CO$_\textnormal{2}$. \all{However, more research is needed to see whether the current method would succeed in these complex areas that exhibit very strong variations of the background wave velocity.} Finally, \all{matrix imaging} can be taken advantage of for marine exploration where the variations of sea local properties can distort the acoustic wave-front depending on the gradients of temperature and salinity, not to mention streams and swirls. The matrix approach presented here can compensate for the phase distortions undergone by the received echoes and improve the interpretation of the data with no more required knowledge than the speed of sound in water. Although not leveraged in the present paper, \cite{william} have also demonstrated the relevance of the matrix approach at removing multiple reverberations between strata layer interfaces. This also may be useful for any marine or seismic exploration purpose.

\section*{ACKNOWLEDGMENTS}
The authors are grateful for funding provided by LABEX WIFI (Laboratory of Excellence within the French Program Investments for the Future, ANR‐10‐LABX‐24 and ANR‐10‐IDEX‐0001‐02 and by TOTAL R\&D. We acknowledge the support from the European Research Council (ERC) under the European Union Horizon 2020 research and innovation program (grant agreement No 742335, F-IMAGE and grant agreement No. 819261, REMINISCENCE). 

\section*{DATA AVAILABILITY}
The seismic data used in this study can be obtained from the Data
Management Center of the Incorporated Research Institutions for
Seismology (IRIS) . The facilities of IRISData Services, and specifically
the IRIS Data Management Center, were used for access to
waveforms, related metadata, and/or derived products used in this
study. IRIS Data Services are funded through the Seismological Facilities
for the Advancement of Geoscience (SAGE) Award of the
National Science Foundation under Cooperative Support Agreement
EAR-1851048.

\vspace{\baselineskip}

\clearpage 

\clearpage

\renewcommand{\thetable}{S\arabic{table}}
\renewcommand{\thefigure}{S\arabic{figure}}
\renewcommand{\theequation}{S\arabic{equation}}
\renewcommand{\thesection}{S\arabic{section}}

\setcounter{equation}{0}
\setcounter{figure}{0}
\setcounter{section}{0}

\begin{center}
\Large{\bf{Supplementary Information}}
\end{center}
\normalsize
This supplementary material includes details about: (\textit{i}) the prior filtering applied to the reflection matrix in the geophone basis; \al{(\textit{ii}) the singular value decomposition of the distortion matrix; (\textit{iii})} the high-order eigenstates of the distortion matrix; \al{(\textit{iv}) the normalized correlation matrix; (\textit{v})} the choice of the wave velocity model based on the focused reflection matrix.\\

\section{Prior filtering of the reflection matrix.}
\begin{figure*}[h!]
 \centering
 \centerline{\includegraphics[width=17cm]{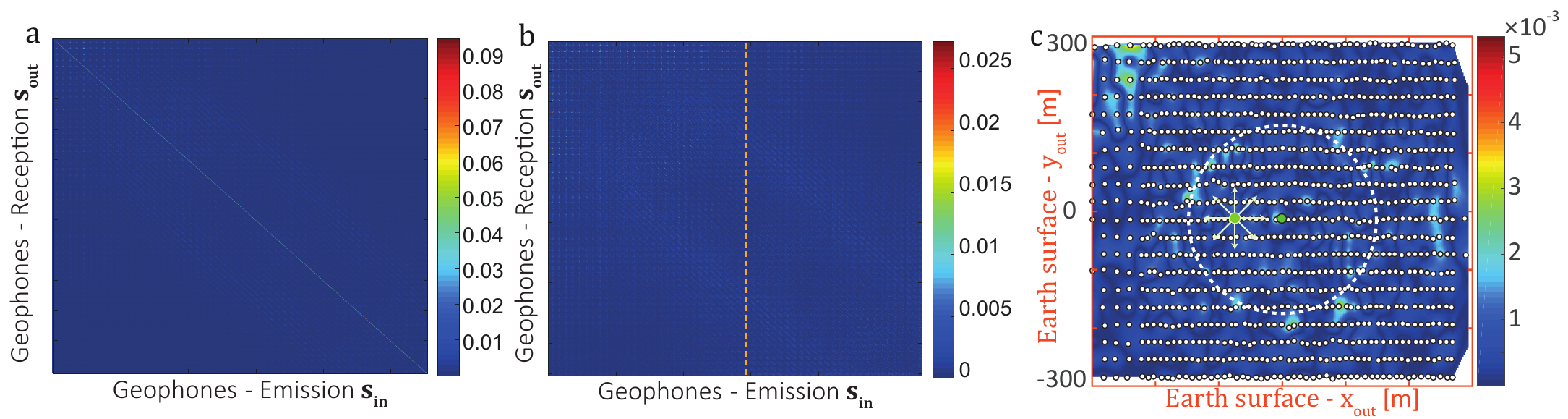}}
 \caption{Reflection matrix in the geophones basis. (a) Raw reflection matrix $\mathbf{K}$ at time $t=0.5$ s. (b) Filtered reflection matrix $\mathbf{K}^\prime$ at time $t=0.5$ s. (c) Reflected wavefield deduced from the spatial interpolation of the central column of $\mathbf{K}^\prime$ ($\mathbf{u}_\textrm{in} = \mathbf{0}$), depicted by a dashed orange line in panel b. The white dashed circle accounts for the expected position of the direct surface wave-front for a Rayleigh wave speed of 350 m/s~\citep{Roux_2016}.}
 \label{figS1}
  \end{figure*}

For the special case of $\mathbf{u}_\textnormal{in}\equiv \mathbf{u}_\textnormal{out}$, the autocorrelation signal $ K(\mathbf{u}_\textnormal{in},\mathbf{u}_\textnormal{in},\tau)$ gives rise to an intense peak at lag time $t=0$, that physically corresponds to the seismic noise power spectral density measured at point $\mathbf{u}_\textnormal{in}$. Only the non zero lag time contribution carries information on the reflectivity at depth and is thus of interest for imaging purposes. However, this zero time peak gathers most of the energy content in the retrieved signal, and can have a detrimental impact on our analysis. In the present situation, the limited frequency bandwidth makes this initial pulse duration far from being negligible ($\delta t \sim 1/\Delta f = $ 0.1 s). It also gives rise to strong side lobes that pollute the relevant signal in the full time range (see Fig.~\ref{figS1}a). Additionally, the ambient seismic wave exhibits a coherence length close to $\lambda/2$. Given the high density of the geophones network, neighbour stations thus belong to the same coherence area. Therefore the corresponding cross-correlation signals, that lie close to the diagonal of $\mathbf{K}$, are also dominated by this autocorrelation peak.

To get rid of this central pulse, a gaussian mask is applied to each element of the raw reflection matrix
\begin{equation*}
    K^\prime ({\mathbf{u}_\textnormal{out}, \mathbf{u}_\textnormal{in}},t) =  K ({\mathbf{u}_\textnormal{out}, \mathbf{u}_\textnormal{in}},t)\times\left[1-\exp \left ({-\frac{\lVert \mathbf{u}_\textnormal{out}-\mathbf{u}_\textnormal{in}\rVert ^2}{ {2\lambda^2}}}\right )\right]
\end{equation*}
where $\lambda$ is the wave-length of body waves at the central frequency (here $\lambda =100$ m). This filter gradually penalizes the impulse responses computed for interstation distances smaller than $\lambda/2$, the coherence length of the body waves in the 10 Hz--20 Hz working frequency range.

Figure \ref{figS1}b displays the filtered version $\mathbf{K}^\prime$ of the measured reflection matrix $\mathbf{K}$ shown in Fig.~\ref{figS1}a. The filtered wave-field is no longer dominated by the seismic noise autocorrelation. Note the strong gap in terms of order of magnitude between the diagonal coefficients of $\mathbf{K}$ (Fig.~\ref{figS1}a) and the actual wave-field patterns in $\mathbf{K}^\prime$ (Fig.~\ref{figS1}b). This pre-filtering operation is thus critical for an accurate interpretation of the effective Earth's response.
	
Each column of $\mathbf{K}^\prime$ corresponds to the wavefield that would be recorded by the set of geophones if a pulse were emitted from one geophone at $\mathbf{u}_\textnormal{in}$. Figure \ref{figS1}c displays the wavefield generated from the central geophone position and interpolated at the surface between all receiving geophones. This wave-field clearly exhibits the  contribution of direct Rayleigh waves that emerges along the white dashed circle. \rita{Interestingly, we observe fast travelling waves (see for instance the top left-hand corner of the array) that correspond to the signature of the body waves propagating below the Earth's surface at larger speed than the surface waves tracked by the geophones.}
They overtake the surface waves and eventually get first to the edge of the array, after reflection on some shallow structures or refraction at overcritical angles.
	
 Figure \ref{figS1}a displays the response matrix $\mathbf{K}(t)$ at time lag $t=0.50$ s. Surprisingly, this matrix is dominated by a predominant signal along its diagonal. To understand the origin of this effect, let us first recall that each coefficient $ K({\mathbf{u}_\textnormal{out}, \mathbf{u}_\textnormal{in}},t)$ of the response matrix is computed from the cross-correlation of the seismic noise wave-field $\psi(\mathbf{u},\tau) $ recorded by geophones located at $\mathbf{u}=\mathbf{u}_\textnormal{out})$ and  $\mathbf{u}=\mathbf{u}_\textnormal{in})$. Under appropriate wave-field conditions, coda cross-correlation converges towards the Green’s function $G({\mathbf{u}_\textnormal{out}, \mathbf{u}_\textnormal{in}},t) $ between receiving stations as if one of them ($\mathbf{u}_\textnormal{in}$) had become a source:
\begin{eqnarray}
    K({\mathbf{u}_\textnormal{out}, \mathbf{u}_\textnormal{in}},t)& = &\lim\limits_{T \rightarrow \infty} \frac{1}{T}\int_{0}^{T} \mathrm{d}\tau \psi(\mathbf{u}_\textnormal{in},\tau)\psi^*(\mathbf{u}_\textnormal{out},t+\tau)\nonumber  \\
    &=&   G({\mathbf{u}_\textnormal{out}, \mathbf{u}_\textnormal{in}},t) - G({\mathbf{u}_\textnormal{out}, \mathbf{u}_\textnormal{in}},-t) 
    \label{SJFZconvergence}
\end{eqnarray}

\section{Singular value decomposition of the distortion matrix}
\label{output}

To decompose the field-of-view into several isoplanatic patches, a SVD of $\Dout$ is performed.  Mathematically, this operation is equivalent to the eigenvalue decomposition of the correlation matrix $\mathbf{C}=\Dout\times\Dout^{\dag}$:
\begin{equation}
   \mathbf{C} =\mathbf{U}_\textrm{out} \times \mathbf{\Sigma}^2 \times \mathbf{U}_\textrm{out} ^\dag
\end{equation}
where the eigenvalues of $ \mathbf{C} $ are the square norm of the singular values $\sigma_i$ of $\Dout$ and its eigenvectors are the output singular vectors $\mathbf{U}^{(i)}_\textrm{out}$. In presence of multiple isoplanatic patches, $\mathbf{C}$ can be decomposed into a set of sub-matrices {$\mathbf{C}^{(p)}$, such that $\mathbf{C}=\sum_p \mathbf{C}^{(p)}$}. Each matrix {$\mathbf{C}^{(p)}$} is associated with a distinct isoplanatic patch $p$ in the field-of-view. In each of them, phase distortions can be modelled by: (\textit{i}) a spatially-invariant input PSF, {$H_\textrm{in}^{(p)}(\mathbf{r},\mathbf{r}')=H_\textrm{in}^{(p)}(\mathbf{r}-\mathbf{r}')$}, in the focused basis; (\textit{ii}) a far-field phase screen of transmittance {$\mathbf{\tilde{H}}_\textrm{out}^{(p)}=[\tilde{H}_\textrm{out}^{(p)}(\kout)]$}.
Using Eq.~(15), the coefficients of {$\mathbf{C}^{(p)}$} can be expressed as follows:{
\begin{equation}
\label{C4}
C^{\alex{(p)}}_\textrm{out} (\mathbf{k},\mathbf{k}') =  \rho^{\alex{(p)}}   \tilde{H}^{\alex{(p)}}_\textrm{out}(\mathbf{k}) \tilde{H}^{(p)*}_\textrm{out}(\mathbf{k}') \left [\tilde{H}^{(p)}_\textrm{in} \circledast \tilde{H}^{(p)}_\textrm{in} \right ](\mathbf{k}-\mathbf{k}'),
\end{equation}
where $\rho^{\alex{(p)}} $ is the overall patch reflectivity and the symbol {$\circledast$ stands for a correlation product}. The correlation function in Eq.~(\ref{C4}) results from the Fourier transform of the scattering distribution {$\left |H^{(p)}_\textrm{in}(\mathbf{r}) \right |^2$} exhibited by the virtual scatterer shown in Fig.~3d. Its support is the correlation width $\Delta k_\textrm{out}$ of the aberration phase transmittance that scales as the inverse of the spatial extension {$\delta_\textrm{in}^{(0)}$} of the input PSF intensity {$\left |H_\textrm{in}^{(p)} \right|^2$}: {$\Delta k_\textrm{out} =\lambda z/\delta_\textrm{in}^{(0)}$}. If the virtual scatterer was point-like, this correlation function would be constant and the matrices {$\Coutp$} would be of rank 1. The corresponding eigenvalue $\sigma_1^2$ would then be equal to $\rho^{{(p)}} $ and the associated eigenvector \alex{$\mathbf{U}_\textrm{out}^{(p,1)}$} would directly yield the aberration transmittance { $\mathbf{\tilde{H}}^{\alex{(p)}}_\textrm{out}$}. \alex{However, i}n practice, the virtual scatterer is of finite size and the correlation term in Eq.~(\ref{C4}) is not negligible. The rank of {$\Coutp$} then scales as the number of \alex{transverse} resolution cells mapping the virtual scatterer~\citep{Robert2009,aubry2006}: {$Q=\left (\delta_\textrm{in}^{(0)} /\delta_0 \right )^2$}. $Q$ is typically equal to $4$ for the local PSF { $\left |H_\textrm{in}^{(p)} \right |^2$} displayed in Fig. 2b. Among these $Q$ eigenmodes of {$\Coutp$}, only the first eigenvector {$\mathbf{U}^{(p,1)}_{\textrm{out}}$} is of interest since it maximizes the backscattered energy by focusing at the center of the virtual scatterer \alex{(see Supplementary Section S3)}.} 

In the following, we thus expect an overall distortion matrix $\Dout$ of rank $P \times Q$ with the following singular value distribution: (\textit{i}) $P$ largest singular values associated with the main eigenstates {$\mathbf{U}^{(p)}_\textrm{out}$} of each patch $p$; (\textit{ii}) a set of $(P-1)\times Q$ of smaller singular values associated with secondary eigenstates that focus on the edges of the virtual scatterers. Figure 4a confirms this prediction by displaying the normalized singular values, {$\hat{\sigma}_{i}={\sigma}_i/\sqrt{\sum_{j=1}^N  {\sigma}_j^2 }$}, of $\mathbf{D}_\textrm{out}$ at depth $z=3600$ m ($\alex{t_B}=4.8$ s). As expected, a few singular values potentially associated with the signal subspace seem to predominate over a continuum of eigenvalues characteristic of the noise subspace. However, it is difficult to determine the effective rank of the signal subspace. This issue can be circumvented by computing the Shannon entropy $\mathcal{H}$ of the singular values~\citep{Badon2019,william}, such that
{
\begin{equation}
 \mathcal{H}= \mich{-}\sum_{i=1}^{N}   \hat{\sigma}_i^2 \log_2 \left [ \hat{\sigma}_i^2 \right ].
   \label{entropy}
\end{equation}}
The Shannon entropy can be used as an indicator of the rank of the signal subspace. In the case of Fig.~4, the singular values of $\mathbf{D}_\textrm{out}$ display an entropy $\mathcal{H}\simeq 9.2$. As the effective rank of $\Dout$ scales as $P \times Q$ and that $Q\sim 4$,  this means that the number $P$ of isoplanatic patches is roughly equal to $2$. \alex{Hence, only the two first eigenstates of $\Dout$ should be considered in the present case. As shown below, the higher-order eigenstates do not bring any supplementary information on aberrations and are shown to be useless for any imaging purpose.}

\section{High-order eigenstates  of the $\mathbf{D}$-matrix.}
  \begin{figure*}
 \centering
 \centerline{\includegraphics[width=8.5cm]{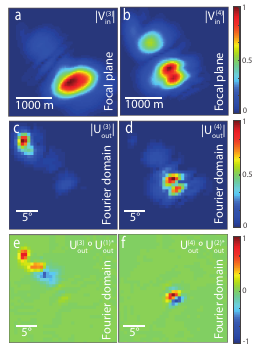}}
 \caption{Singular value decomposition of the distortion matrix $\Dout$ at time $t_B=4.8$ s and depth $z=$3600 m. (a,b) Modulus of input eigenvectors {$\mathbf{V}^{(3)}_{\textrm{in}}$ and $\mathbf{V}^{(4)}_{\textrm{in}}$}. (d,e) Modulus of output eigenvectors {$\mathbf{U}^{(3)}_{\textrm{out}}$ and $\mathbf{U}^{(4)}_{\textrm{out}}$}. (e) Hadamard product between {$\mathbf{U}^{(3)}_{\textrm{out}}$ and $\mathbf{U}^{(1)*}_{\textrm{out}}$}. (f) Hadamard product between {$\mathbf{U}^{(4)}_{\textrm{out}}$ and $\mathbf{U}^{(2)*}_{\textrm{out}}$}.}
 \label{figS2}
  \end{figure*}
In this section, the higher-order eigenstates of $\Dout$ are investigated theoretically and shown to be higher-order modes of the two first ones. 

In Fig.4a of the accompanying paper, the SVD of $\Dout$ shows four main singular values on top of a ``noise'' continuum. However, only the two first eigenstates (Fig.~5) are then used to compensate for aberrations over two main isoplanatic patches. Fig.~\ref{figS2} displays the two following eigenstates of rank 3 and 4. While the third input singular vector, $\mathbf{V}^{(3)}_{\textrm{in}}$ (Fig.~\ref{figS2}a), seems associated with the first isoplanatic patch (see the comparison with $\mathbf{V}^{(1)}_{\textrm{in}}$ in  Fig.~4b), the fourth one, $\mathbf{V}^{(4)}_{\textrm{in}}$ (Fig.~\ref{figS2}b), is also linked to the second isoplanatic patch (see the comparison with $\mathbf{V}^{(2)}_{\textrm{in}}$ in Fig.~4c). Similarly, the ouptput singular vectors $\mathbf{U}^{(3)}_{\textrm{out}}$ (Fig.~\ref{figS2}c) and $\mathbf{U}^{(4)}_{\textrm{out}}$ (Fig.~\ref{figS2}d) map onto the same angular domain as $\mathbf{U}^{(1)}_{\textrm{out}}$ (Fig.~4d) and $\mathbf{U}^{(2)}_{\textrm{out}}$ (Fig.~4e), respectively. In this Supplementary Section, we explain how these higher-order eigenstates of $\Dout$ are linked to the first-order ones and why they are not useful for an imaging purpose. 

To that aim, we will first study theoretically the SVD of $\Doutp$ over a single isoplanatic patch $p$ before extending those predictions to the general case of transversely-varying aberrations. Over each isoplanatic patch $p$, the corresponding correlation matrix, $\Coutp=\Doutp \times \mathbf{D}_\textrm{out}^{(p){\dag}}$, is equivalent to a reflection matrix for a virtual reflector of characteristic size $\delta_{\textrm{in}}^{(0)}$. In such configuration, the rank of $\Coutp$ scales as the number of resolution cells, $Q=\left ( \delta_{\textrm{in}}^{(0)}  / \delta_0 \right )^2$, mapping the virtual reflector~\citep{Robert2009}. The corresponding eigenvectors $\mathbf{U}_{\textrm{out}}^{(i)}$ can be expressed as the Hadamard product between the far-field aberration phase law $\mathbf{H}_{\textrm{out}}^{(p)}$ and eigenmodes $\mathbf{W}_{\textrm{in}}^{(i)}$ of the correlation kernel,  $\mathbf{\tilde{H}}_{\textrm{in}}^{(p)} \circledast \mathbf{\tilde{H}}_{\textrm{in}}^{(p)} =[\tilde{H}_{\textrm{in}}^{(p)} \circledast  \tilde{H}_{\textrm{in}}^{(p)} (\mathbf{k}-\mathbf{k}')]$, such that
\begin{equation}
\mathbf{U}_{\textrm{out}}^{(i)} = \mathbf{\tilde{H}}_{\textrm{out}}^{(p)}  \circ \mathbf{W}_\textrm{in}^{(i)}
\end{equation}
The shape of the eigenmodes $\mathbf{W}_\textrm{in}^{(i)}$ depends on the exact form of the correlation function $\tilde{H}_{\textrm{in}}^{(p)} \circledast  \tilde{H}_{\textrm{in}}^{(p)}$. For instance, a sinc correlation function imply 3D prolate spheroidal eigenfunctions~\citep{Robert2009}; a Gaussian covariance function leads to Hermite-Gaussian eigenmodes~\citep{aubry2006}. The first eigenmode is given, in first approximation, by~\citep{Badon2019}:
\begin{equation}
\mathbf{W}_\textrm{in}^{(1)}(\mathbf{k})=\tilde{H}_{\textrm{in}}^{(p)} \circledast  \tilde{H}_{\textrm{in}}^{(p)}(\mathbf{k})
\end{equation}
As the correlation function $\tilde{H}_{\textrm{in}}^{(p)} \circledast  \tilde{H}_{\textrm{in}}^{(p)}(\mathbf{k})$ is, in first approximation, real and positive, a general trend is that the first eigenmode ${W}_\textrm{in}^{(1)}(\mathbf{k})$ displays a nearly constant phase. This is a very important property since it means that the phase of the first eigenvector $\mathbf{U}_{\textrm{out}}^{(1)}$ is a direct estimator of $\mathbf{\tilde{H}}_{\textrm{out}}^{(p)}$. The higher rank eigenvectors $W_\textrm{in}^{(i)}$ are more complex and exhibit a number of lobes that scales with their rank $i$. The corresponding eigenvectors $\mathbf{U}_{\textrm{out}}^{(i)}$ do not bring a priori any useful information compared to the fundamental one $\mathbf{U}_{\textrm{out}}^{(1)}$  as their far-field phase law is modulated by eigenfunctions ${W}_\textrm{in}^{(i)}(\mathbf{k})$ that display a number $(i-1)$ of sign reversals. As a consequence, they cannot be used for any imaging purpose since the associated PSF will be made of $i$ lobes~\citep{Robert2009,aubry2006}. On the contrary the PSF associated with $\mathbf{U}_{\textrm{out}}^{(1)}$ shows a single central lobe.

In the general case (several isoplanatic patches in the field-of-view), the rank of $\Dout$ and $\Cout$  scales as $P \times Q$, with $P$ the number of isoplanatic patches. By estimating this rank from the Shannon entropy of the singular values (Eq.~19) and the number $Q$ from the imaging PSF (Fig.~2b), we can estimate the number of significant isoplanatic patches (here roughly equal to 2). We thus expect that the two first eigenstates of $\Dout$  and $\Cout$ are associated with distinct isoplanatic patches and that higher-rank eigenstates are just higher-order eigenmodes derived from the two fundamental eigenstates. 

Nevertheless, from the singular value spectrum displayed in Fig.~4a, it is difficult to state that the third and fourth eigenstates should not be considered. Their singular values clearly emerge from the continuum of lowest singular values that can be seen as a ``noise'' background. A first check consists in looking at the input eigenvectors $\mathbf{V}_{\textrm{in}}^{(i)}$ that are supposed to map onto the corresponding isoplanatic patches. Fig.~4b and c indeed show that the two first eigenstates map onto two disjoint areas. This observation proves that the two first eigenstates are associated with different isoplanatic patches. On the contrary, the third and fourth eigenstates, $\mathbf{V}_{\textrm{in}}^{(3)}$ and $\mathbf{V}_{\textrm{in}}^{(4)}$, are shown to focus on the same areas but with a more complex pattern (see Fig.~\ref{figS3}a and b, respectively). To show the link between these high-order eigenstates and the first-order ones, one can investigate the associated output eigenvectors $\mathbf{U}_{\textrm{out}}^{(i)}$ (see Fig.~\ref{figS3}c and d). Indeed, if they are associated with the same aberration phase law $\mathbf{\tilde{H}}_{\textrm{out}}^{(p)}$, this product should be real and exhibits a sign reversal characteristic of the second order mode ${W}_\textrm{in}^{(2)}(\mathbf{k})$. The link between the first and third eigenstates is confirmed by Fig.~\ref{figS3}e that displays the real part of the Hadamard product $\left[ \mathbf{U}_{\textrm{out}}^{(3)}  \circ \mathbf{U}_{\textrm{out}}^{(1)*} \right]$. A sign reversal characteristic of ${W}_\textrm{in}^{(2)}(\mathbf{k})$ is thus revealed. A similar link between $\mathbf{U}^{(2)}_{\textrm{out}}$ and $\mathbf{U}^{(4)}_{\textrm{out}}$ is demonstrated by  Fig.~\ref{figS3}f that displays the product  $\left[ \mathbf{U}_{\textrm{out}}^{(4)}  \circ \mathbf{U}_{\textrm{out}}^{(2)*} \right]$. 

This detailed analysis confirms that the third and fourth eigenstates correspond to second-order modes of the first ones. Hence they do not bring any supplementary information on subsoil-induced aberrations, hence they should not be considered for matrix imaging. Beyond this example at this specific time-of-flight, it should be noted that the same behavior is observed over the whole depth range. Two main isoplanatic patches dominate the singular value spectrum. The fault structure probably explains this peculiar behavior: The two different wave velocity distribution on each side of the fault implies the presence of two main isoplanatic patches in depth.

\section{Normalized correlation matrix}

\al{In the last part of the matrix imaging process (Section 4.3), an output residual distortion matrix $\delta \mathbf{D}^{(p)}_\textrm{out}$ is considered. By analogy with Eq.~\ref{C4}, the corresponding correlation matrix, {$\delta \mathbf{C}_\textrm{out}^{(p)}=\delta\mathbf{D}^{(p)}_\textrm{out} \times \delta\mathbf{D}_\textrm{out}^{(p)\dag}$}, can be expressed as follows:}{
\begin{equation}
\label{C5}
\delta C_\textrm{out}^{(p)} (\kout,\kpout) \propto  \rho_p \delta \tilde{H}_\textrm{out}^{(p)}(\kout) \delta \tilde{H}_\textrm{out}^{(p)*}(\kpout)  \left [\delta \tilde{H}_\textrm{in}^{(p)} \circledast \delta \tilde{H}_\textrm{in}^{(p)} \right ](\kout-\kpout).
\end{equation}}
As previously highlighted in Sec.~\ref{output} [Eq.~(\ref{C4})], the correlation term in Eq.~(\ref{C5}) prevents a proper estimation of the aberration phase transmittance over the whole angular spectrum. To circumvent that issue, the correlation matrix coefficients can be normalized [Eq.~(25)]. As illustrated by Fig.~3e, this operation makes the virtual reflector point-like~\citep{william}. Indeed, if we make the realistic hypothesis of a real and positive autocorrelation function {$\left [ \delta \tilde{H}_\textrm{in}^{(p)} \circledast \delta \tilde{H}_\textrm{in}^{(p)}\right ]$ in Eq.~\eqref{C5}, the coefficients of $\delta \mathbf{C}^{(p)}_\textrm{out}$} can actually be expressed as follows:{
\begin{equation}
\label{C6}
\delta C^{(p)}_\textrm{out}(\kout,\kpout) \propto   \delta \tilde{H}^{(p)}_\textrm{out}(\kout) \delta \tilde{H}^{(p)*}_\textrm{out}(\kpout) .
\end{equation}}
Such a matrix is equivalent to the time reversal operator associated with a point-like reflector at the origin~\citep{prada,Prada1996}. In that case, the matrix {$\delta \mathbf{\hat{C}}^{(p)}_\textrm{out}$} is of rank 1 and its eigenvector {$\delta \mathbf{U}^{(p)}_\textrm{out}$} yields the residual aberration phase transmittance:{
\begin{equation}
\delta \mathbf{U}^{(p)}_\textrm{out} (\kout)= \delta \tilde{H}^{(p)}_\textrm{out}(\kout).
\end{equation}}
The phase conjugate of $\delta \mathbf{U}^{(p)}_\textrm{out}$ yields the additional focusing law that is applied to the reflection matrix in order to overcome the residual output aberrations [Eq. (27)].

\section{Choice of the wave velocity model.}
  \begin{figure*}
 \centering
 \centerline{\includegraphics[width=17cm]{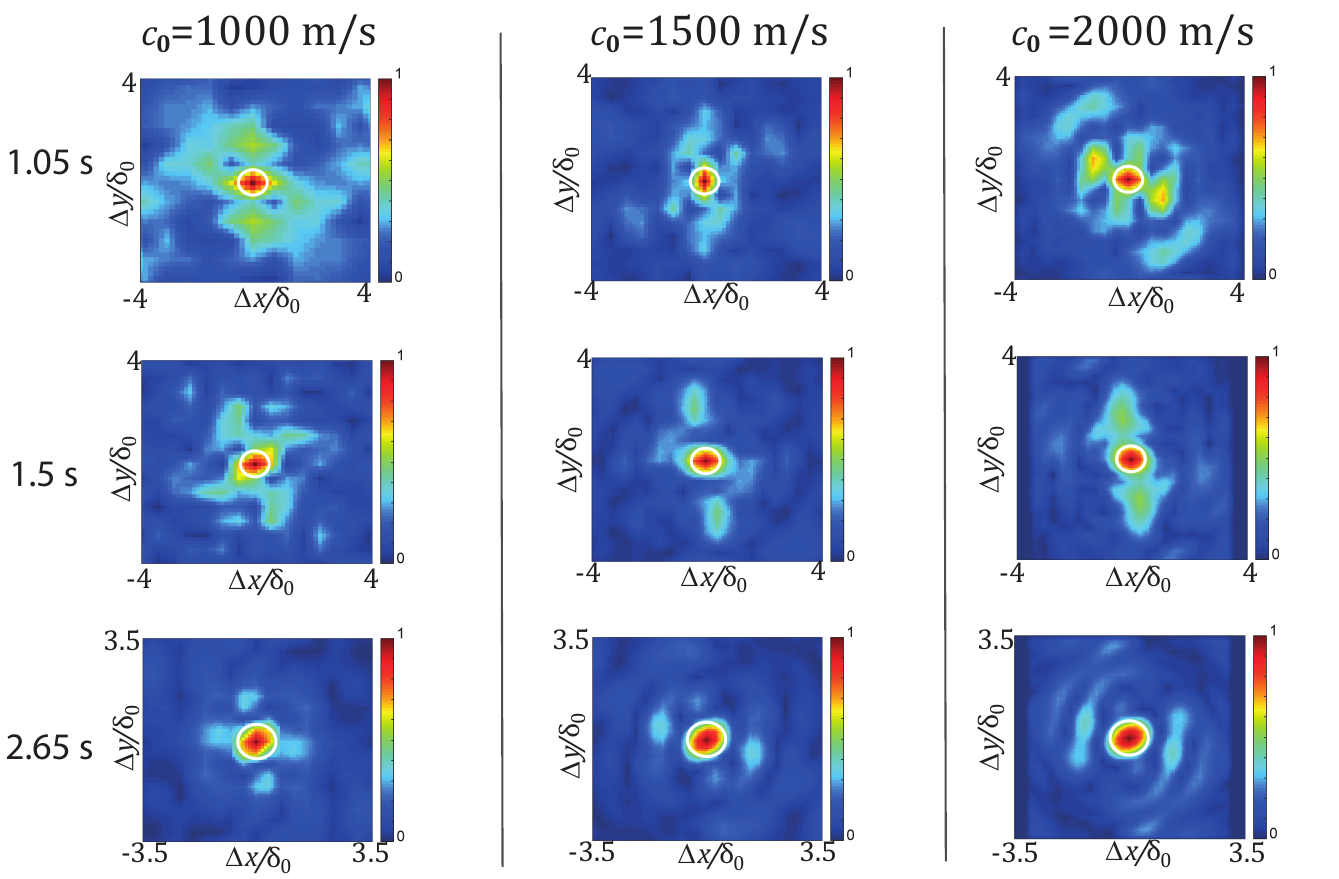}}
 \caption{Imaging PSF deduced from the antidiagonal of $\mathbf{R}$ whose common mid-point exhibits the maximum confocal signal. This imaaging PSF is shown at different times of flight ($t_B=1.05$, 1.5 and 2.65 s, from top to bottom) for different wave velocity models: $c_0=1000$ m/s (a), $c_0=1500$ m/s (b), $c_0=2000$ m/s (c). Each panel displays the modulus of the reflected wave-field normalized by its maximum. The white circle accounts for the theoretical resolution cell (disk of radius $\delta_0$) imposed by the geophone array aperture.}
 \label{figS3}
  \end{figure*}
In the accompanying paper, the reflected body waves are used to build a high-resolution in-depth image of the SJFZ. To that aim, the choice of the initial wave velocity model is crucial. As mentioned in the accompanying paper, this choice can be enlightened by the properties of the focused reflection matrix $\mathbf{R}$. In particular, a common mid-point intensity profile can be extracted from the main antidiagonal of $\mathbf{R}$ that exhibits the maximum confocal (diagonal) signal. Figure~\ref{figS3} shows the corresponding intensity profile at several times of flight ($t_B=1.05$, 1.5 and 2.65 s, from top to bottom) and for different seismic wave velocity ($c_0=1000$, 1500 and 2000 m/s, from left to right) in our propagation model. To make the comparison quantitative, each intensity profile is displayed as a function of spatial coordinates normalized by $\delta_0$, the expected resolution for each velocity model at the corresponding depth. In each panel, the focal spot consists of a main lobe centered around the input focusing point and a random distribution of secondary lobes. The dimension of the main lobe is roughly twice the diffraction limit prediction $\delta_0$, depicted by a white circle in each panel Fig.~\ref{figS3}. This loss of resolution is a manifestation of the aberrations induced by the gap between the velocity model and the actual seismic wave velocity distribution in the SJFZ underground. However, the dimension of this main lobe is not significantly affected by our choice of $c_0$. Hence this observable cannot be used for optimizing our wave propagation model. On the contrary, the amplitude and spatial extension of the secondary lobes strongly depend on the wave velocity model. The value $c_0=1500$ m/s is the seismic wave velocity that clearly minimizes the level of these secondary lobes in Fig.~\ref{figS3}. 

\begin{figure*}
 \centering
 \centerline{\includegraphics[width=17cm]{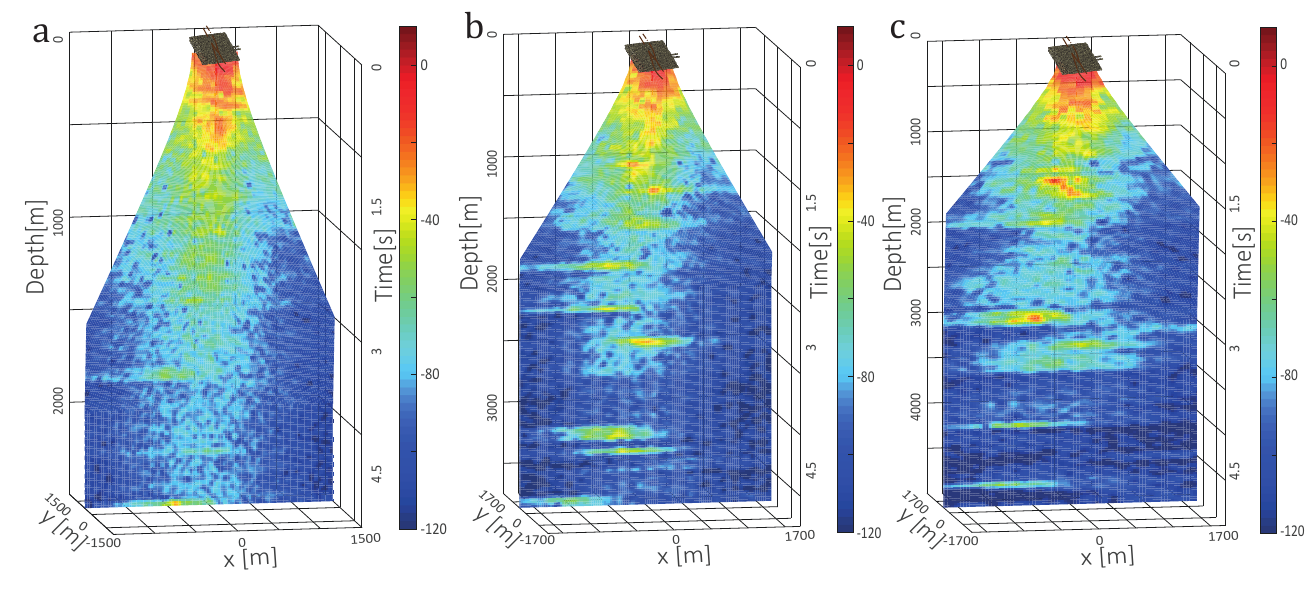}}
 \caption{Vertical slice of the 3D matrix images computed under the matrix approach for different seismic wave velocity models:  $c_0=1000$ m/s (a), $c_0=1500$ m/s (b), $c_0=2000$ m/s (c). The slice orientation is chosen to be normal to the fault plane. The color scale for each image is in dB.}
 \label{figS4}
  \end{figure*}

The choice is validated \textit{a posteriori} by the quality of the confocal image obtained under our matrix approach. Figure \ref{figS4} shows a slice of this image for different wave velocity models ($c_0=1000$, 1500 and 2000 m/s). The slice orientation is the same as the b-scans displayed in the accompanying paper. While the raw image $\mathcal{I}_0$ is completely blurred (Fig.~2c of the accompanying paper), a gain in resolution is provided by the matrix aberration correction process. However, the image quality strongly depends on the choice of the wave velocity model. This sensitivity can be explained by the fact that axial aberrations are not tackled by our matrix approach so far. Hence, an optimized wave velocity model allows to properly capture all the echoes back-scattered at the focal depth $z$ in $\mathbf{R}(z)$. The comparison between Fig.\ref{figS4}a and b confirms that a model with $c_0=1000$ m/s clearly underestimates the actual body wave velocity in the depth range considered in this work. While Fig.~\ref{figS4}b ($c_0=1500$ m/s) clearly highlights strata layers at different depths on both sides of the fault, Fig.~\ref{figS4}a ($c_0=1000$ m/s) shows a blurred image of the subsoil. For $c_0=2000$ m/s (Fig.\ref{figS4}c), the strata structure of the SJFZ underground is partially revealed but with a worse resolution and lower contrast than for $c_0=1500$ m/s (Fig.\ref{figS4}b) until an echo time $t$=4 s. Nevertheless, note that, beyond that time, the two last interfaces seem to be better resolved for $c_0=2000$ m/s. This indicates that, not surprisingly, the seismic wave velocity increases with depth and that $c_0=1500$ m/s is probably not the optimal wave velocity for $t>4$ s. A mapping of the wave velocity could be indeed possible through the focused reflection matrix approach as already demonstrated in ultrasound imaging~\citep{Lambert2020}. However, the extension of this method to seismology is beyond the scope of this paper. The matrix mapping of the bulk seismic wave velocity will be tackled in future works.
	
%

\end{document}